\documentclass[a4paper,12pt]{article}
\usepackage[utf8]{inputenc}

\usepackage[margin=2.5cm]{geometry}
\usepackage{graphicx}
\usepackage{caption}

\usepackage{subfig}

\usepackage{hyperref}

\usepackage{amsmath}

\usepackage{color}

\usepackage{authblk}

\title{Quasinormal modes of Kerr black holes using a spectral decomposition of the metric perturbations}

\author[1]{Jose Luis Bl\'azquez-Salcedo\thanks{\href{mailto:jlblaz01@ucm.es}{jlblaz01@ucm.es}}}
\author[2]{Fech Scen Khoo\thanks{\href{mailto:fech.scen.khoo@uni-oldenburg.de}{fech.scen.khoo@uni-oldenburg.de}}} 
\author[2]{Jutta Kunz\thanks{\href{mailto:jutta.kunz@uni-oldenburg.de}{jutta.kunz@uni-oldenburg.de}}} 
\author[1]{Luis Manuel Gonz\'alez-Romero\thanks{\href{mailto:mgromero@ucm.es}{mgromero@ucm.es}}} 
\affil[1]{Departamento de F\'isica Te\'orica and IPARCOS, Facultad de Ciencias F\'isicas, Universidad Complutense de Madrid, Spain}
\affil[2]{Institut f\"ur  Physik, Universit\"at Oldenburg, Postfach 2503,
D-26111 Oldenburg, Germany}

\date{\today}

\begin{document}

\maketitle

\begin{abstract}
We report a new method to calculate the quasinormal modes of rotating black holes, using a spectral decomposition to solve the partial differential equations that result from introducing linear metric perturbations to a rotating background.
Our approach allows us to calculate a large sector of the quasinormal mode spectrum. In particular, we study the accuracy of the method for the $(l=2)$-led and $(l=3)$-led modes for different values of the $M_z$ azimuthal number, considering the fundamental modes as well as the first two excitations. We show that our method reproduces the Kerr fundamental modes with an accuracy of $10^{-6}$ or better for $a/M<0.8$, while it stays below $0.1\%$ for $a/M<0.98$.
\end{abstract}

\section{Introduction}

The observation of gravitational waves from the merging of black holes and neutron stars has given us a new window to the Universe and, in particular, a powerful new means to learn about the strong gravity sector
\cite{LIGOScientific:2016aoc,LIGOScientific:2017vwq}.
Future gravitational wave detectors on the ground and in space will allow us to scrutinize the numerous alternative gravity theories, that have been proposed, by investigating their predictions for the merger of compact objects \cite{Berti:2015itd,Barack:2018yly,Maggiore:2019uih,CANTATA:2021ktz}.

The Schwarzschild and Kerr black holes of General Relativity are rather special objects, since these spacetimes are fully characterized by only two quantities, their mass and their angular momentum \cite{Chrusciel:2012jk}.
But in alternative gravity theories black holes may possess more distinctive properties, i.e., they may carry hair (see e.g., \cite{Herdeiro:2015waa,Volkov:2016ehx}).
Besides, even in General Relativity hairy black holes of potential astrophysical interest may arise when fields beyond the Standard Model are considered \cite{Herdeiro:2014goa,Herdeiro:2016tmi}.

While the merging of black holes proceeds via inspiral, merger and ringdown, a full analysis of all these steps in Numerical Relativity would be too costly, in particular, in view of the many possible parameter combinations and (still) viable alternative gravity theories.
A more efficient way to constrain or rule out alternative gravity theories could be to focus on a subset of the gravitational waves emitted in black hole mergers.
Therefore our goal is to determine the quasinormal modes emitted in the ringdown after the merger of black holes for a variety of alternative gravity theories and their rapidly rotating black holes.

During the ringdown the newly formed excited black hole emits gravitational waves to finally settle down to its stationary limit.
The quasinormal modes are thus obtained by studying black hole perturbations.
The complex eigenvalues associated with quasinormal modes consist of the real part describing the frequency of the oscillation and the imaginary part yielding the decay time.
The quasinormal modes of the Schwarzschild and Kerr black holes are well known (see e.g., \cite{Kokkotas:1999bd,Berti:2009kk,Konoplya:2011qq}).
But already the presence of the electromagnetic field has posed a challenge to obtain the quasinormal modes of the rotating Kerr-Newman black holes, which was accomplished only a few years ago using numerical methods \cite{Dias:2015wqa,Dias:2021yju,Dias:2022oqm} (see also \cite{Pani:2013ija,Pani:2013wsa,Blazquez-Salcedo:2022eik} for perturbative studies).

The studies of quasinormal modes of black holes in alternative gravity theories have focused so far on static spherically symmetric black holes.
These include, for instance, quasinormal modes of Schwarzschild black holes in some Horndeski theories \cite{Tattersall:2017erk,Tattersall:2018nve,Tattersall:2019nmh} and
in dynamical Chern-Simons theories \cite{Cardoso:2009pk,Molina:2010fb,Kimura:2018nxk}, or
the quasinormal modes of hairy black holes in Einstein-dilaton-Gauss-Bonnet theories \cite{Kanti:1997br,Pani:2009wy,Ayzenberg:2013wua,Blazquez-Salcedo:2016enn,Blazquez-Salcedo:2017txk,Blazquez-Salcedo:2018pxo,Konoplya:2019hml,Zinhailo:2019rwd} and Einstein-scalar-Gauss-Bonnet theories \cite{Blazquez-Salcedo:2018jnn,Silva:2018qhn,Macedo:2020tbm,Blazquez-Salcedo:2020rhf,Blazquez-Salcedo:2020caw}.

{ 
In recent years, though, also strategies for obtaining quasinormal modes of rotating hairy black holes have been pushed forward \cite{Li:2022pcy,Cano:2023tmv}.
In particular, perturbative studies of quasinormal modes of slowly rotating black holes have been performed in higher-derivative gravity \cite{Cano:2020cao, Cano:2021myl,Cano:2023jbk}, } 
Einstein-dilaton-Gauss-Bonnet theory \cite{Pierini:2021jxd,Pierini:2022eim}, Chern-Simons theory \cite{Wagle:2021tam,Wagle:2023fwl}, and Einstein-bumblebee gravity \cite{Liu:2022dcn}.
{Moreover, quasinormal modes for test fields in the background of rapidly rotating black holes have been studied, as well as gravitational perturbations employing an ad hoc deformation of the wave equation \cite{Ghosh:2023etd, Konoplya:2022pbc}.}

The true challenge, however, still remains for the study of the full set of quasinormal modes of rapidly rotating black holes in alternative gravity theories. %
In order to develop and test appropriate methods to tackle this challenge, we have considered a scheme based on a spectral decomposition of the metric perturbations.
We have been inspired by the successful application of such methods to the quasinormal modes of neutron stars \cite{Kruger:2019zuz,Kruger:2020ykw,Kruger:2021zta}.
We note, that a spectral scheme has also been applied successfully to the Schwarzschild case earlier this year \cite{Chung:2023zdq}.

Here we outline our numerical scheme and present first results testing our scheme with the well-known quasinormal modes of Kerr black holes.
In section \ref{setup} we provide the general setting with the field equations and the Kerr solution.
We discuss the metric perturbations in section \ref{met_pert}, where we present the Ansatz, the perturbation equations, our parametrization, the boundary conditions and the spectral decomposition.
Section \ref{results} shows our results for the scalar modes and the metric modes. We end with our conclusions in section \ref{conclusions}.

\section{General setting}
\label{setup}

\subsection{Theory and field equations}

We consider the standard action
of an Einstein-Hilbert term
with a minimally coupled scalar field $\varphi$,

\begin{eqnarray}
			S&=&\frac{1}{2\kappa}\int d^4x \sqrt{-g} 
		\Big[R - \frac{1}{2} \partial_\mu \varphi \, \partial^\mu \varphi 
		 \Big] \, .
   \label{eq:quadratic} 
\end{eqnarray}
The Einstein equations are
\begin{eqnarray}
\mathcal{G}_{\mu\nu} = G_{\mu\nu} - T_{\mu\nu} = 0 \, ,
\end{eqnarray}
where $G_{\mu\nu}$ is the Einstein tensor and $T_{\mu\nu}$ is the scalar stress energy tensor, which is given by
\begin{eqnarray}
T_{\mu\nu} =  \frac{1}{2}\partial_{\mu}\varphi\,\partial_{\nu}\varphi - \frac{1}{4}g_{\mu\nu}  (\partial \varphi)^2  \, .
\end{eqnarray}
On top of that, we have the scalar field equation,
\begin{eqnarray}
 \mathcal{S} =\frac{1}{\sqrt{-g}} \partial_{\mu} (\sqrt{-g} g^{\mu\nu} \partial_{\nu} \varphi)  = 0 \, .
\end{eqnarray}

\subsection{Kerr solution in Boyer-Lindquist coordinates}

In Boyer-Lindquist coordinates $({r},{\theta})$, 
the Kerr solution is
\begin{eqnarray}
ds^2 = 
&-& \left(1-\frac{2Mr}{r^2+a^2\cos^2\theta}\right) dt^2  
+ \left(\frac{r^2+a^2\cos^2\theta}{r^2+a^2-2Mr}\right) dr^2  
+ \left(r^2+a^2\cos^2\theta \right) d\theta^2  \nonumber \\
&+& \left(r^2+a^2+\frac{2Ma^2r\sin^2\theta}{r^2+a^2\cos^2\theta}\right) d\phi^2  - \frac{4Mar\sin^2\theta}{r^2+a^2\cos^2\theta} dt d\phi
\, ,
\label{metric_1}
\end{eqnarray}
where $a$ is the Kerr parameter, that is related to the angular momentum $J$ and the black hole mass $M$: $a=J/M$. The black hole horizons are located at
\begin{equation}
   {r}_{\pm} = M \pm \sqrt{M^2-a^2} \, .
\end{equation}
Here we will work with the outer horizon, $r_+\equiv r_H$. 
The exterior part of the black hole solution is defined in the domain of $r\in [r_H,\infty)$, $\theta \in [0,\pi]$, and $\phi \in [0,2\pi)$.

The area of the black hole horizon is
\begin{eqnarray}
A_H = 4\pi (r_H^2 + a^2) \, ,
\end{eqnarray}
and the horizon angular velocity is given by
\begin{eqnarray}
\Omega_H = \frac{a}{r^2_H + a^2}\, .
\end{eqnarray}
In the extremal Kerr limit
$r_+=r_-$, the following relation
\begin{eqnarray}
M=|a| \, 
\end{eqnarray}
holds between the black hole mass and the Kerr parameter.

\section{General metric perturbations of %
Kerr
}
\label{met_pert}

\subsection{Ansatz}
\label{sec_ansatz}

Here we perturb Kerr black holes non-radially.
The perturbations are tracked to linear order by the auxiliary parameter $\epsilon$.
The background metric $g^{(bg)}_{\mu\nu}$ is given by equation (\ref{metric_1}),
while the background scalar field $\varphi$ vanishes. The superscript $(bg)$ denotes the background. 

We keep the perturbations as general as possible, and decompose only the time dependence into harmonics. 
Additionally we make use of the axial symmetry of the background, and factorize the $\phi$-dependence of the perturbations with the corresponding harmonic dependence, 
namely by introducing the azimuthal number $M_z$.

The full metric can be written like
\begin{eqnarray}
g_{\mu\nu} &=& g^{(bg)}_{\mu\nu} + \epsilon \delta h_{\mu\nu}(t,r,\theta,\phi) \, , 
\end{eqnarray}
and the metric perturbations can be further separated into axial and polar components, 
\begin{eqnarray}
\delta h_{\mu\nu} &=& \delta h^{(A)}_{\mu\nu} + \delta h^{(P)}_{\mu\nu}  
\, .
\end{eqnarray}
The superscripts $(A)$ and 
$(P)$ denote axial-led and polar-led perturbations, respectively.
The Ansatz for the axial and polar metric perturbations is, respectively, given by
\begin{equation}
 \delta h^{(A)}_{\mu\nu} = e^{i(M_z\phi-\omega t)} 
\begin{pmatrix}
0             & 0             & a_1(r,\theta) & a_2(r,\theta) \\
0             & 0             & a_3(r,\theta) & a_4(r,\theta) \\
a_1(r,\theta) & a_3(r,\theta) & 0             & 0 \\
a_2(r,\theta) & a_4(r,\theta) & 0             & 0
\end{pmatrix}   
\end{equation}
and
\begin{equation}
 \delta h^{(P)}_{\mu\nu} = e^{i(M_z\phi-\omega t)} 
\begin{pmatrix}
N_0(r,\theta) & H_1(r,\theta) & 0             & 0 \\
H_1(r,\theta) & L_0(r,\theta) & 0             & 0 \\
0             & 0             & T_0(r,\theta) & 0  \\
0             & 0             & 0             & S_0(r,\theta) 
\end{pmatrix}   \, ,
\end{equation}
where the following definitions are convenient in order to fix the gauge and simplify the equations:
\begin{eqnarray}
a_1(r,\theta) &=& - i M_z \frac{h_0(r,\theta)}{\sin{\theta}} \, , \\
a_2(r,\theta) &=& \sin{\theta} \, \partial_\theta h_0(r,\theta) \, ,  \\
a_3(r,\theta) &=& - i M_z \frac{h_1(r,\theta)}{\sin{\theta}} \, ,  \\
a_4(r,\theta) &=& \sin{\theta} \, \partial_\theta h_1 \, ,  \\
N_0(r,\theta) &=& \left( g^{(bg)}_{rr}(r,\theta) \right)^{-1}  N(r,\theta) \, ,  \\
L_0(r,\theta) &=& \left( g^{(bg)}_{rr}(r,\theta) \right)  L(r,\theta) \, ,  \\
T_0(r,\theta) &=& \left( g^{(bg)}_{\theta\theta}(r,\theta) \right) T(r,\theta)  \, , \\
S_0(r,\theta) &=& \left( g^{(bg)}_{\phi\phi}(r,\theta) \right) T(r,\theta) \, .
\end{eqnarray}

In addition we have the perturbation of the scalar field, which is just a test field in the Kerr background,
\begin{eqnarray}
\varphi &=& \varphi^{(bg)} + \epsilon \delta\varphi(t,r,\theta,\phi) =  \epsilon e^{i(M_z\phi-\omega t)} \Phi(r,\theta) \, .
\end{eqnarray}

\subsection{Perturbation equations}

In general, the system of 
equations is described by
\begin{eqnarray}
\mathcal{G}_{\mu\nu} = \mathcal{G}_{\mu\nu}^{(bg)} + \epsilon \delta\mathcal{G}_{\mu\nu} (r,\theta) e^{i(M_z\phi-\omega t)}  =0 \, , \\
\mathcal{S} = \mathcal{S}^{(bg)} + \epsilon \delta\mathcal{S} (r,\theta) e^{i(M_z\phi-\omega t)}   =0 \, .
\end{eqnarray}
Since the background is Kerr, the equations $\mathcal{G}_{\mu\nu}^{(bg)}$ and $\mathcal{S}^{(bg)}$ are all trivially zero.
The components $\delta\mathcal{G}_{\mu\nu}$ result in a system of partial differential equations (PDEs)
in $r$ and $\theta$ for
the metric perturbations,
while
$\delta\mathcal{S}$ is a 
PDE 
for the scalar field perturbation $\Phi(r,\theta)$. The equations are linear in the perturbation functions, 
and the coefficients of the linear equations are functions that depend on the background metric components, 
i.e.,~the coefficients are functions of the variables $(r, \theta)$, and the background parameters
such as mass $M$ and Kerr parameter $a$. 
These coefficients also depend on the angular number $M_z$ of the perturbations, and the eigenfrequency $\omega$. 
For example, the scalar equation in the Kerr background is 
\begin{eqnarray}
&&
2(r-M)\frac{\partial \Phi}{\partial r}
+ \frac{\cos\theta}{\sin\theta}\frac{\partial \Phi}{\partial \theta}
+ (r^2+a^2-2Mr) \frac{\partial^2 \Phi}{\partial r^2}
+ \frac{\partial^2\Phi}{\partial \theta^2} 
+ M_z^2 \frac{2Mr-r^2-a^2\cos^2\theta}{(2Mr-a^2-r^2)\sin^2\theta} \Phi 
\nonumber \\
&&+ \omega^2 \frac{\cos^2\theta (2Ma^2r-a^4-a^2r^2) - 2Ma^2r - a^2r^2 - r^4}{2Mr-a^2-r^2} \Phi
+  \frac{4 \omega M_z Ma r}{2Mr-a^2-r^2} \Phi
= 0
\, .
\label{scalar_eq}
\end{eqnarray}
The system of equations resulting from the Einstein equation has a similar structure but is much more involved, since the axial and polar functions couple with each other non-trivially. 

\subsection{Parametrization and equations}

In order to solve the above system of equations and extract the quasinormal mode eigenvalues 
$\omega$, it is convenient to change the parametrization in the following way.
First we choose a
compactification of the coordinates,
\begin{eqnarray}
    x = 1 - r_H/r \, , \, \, \,   y = \cos\theta \, .
\end{eqnarray}
The domain of integration is then
$0 \le x \le 1$ and $-1 \le y \le 1$. The horizon is located at
$x=0$, and at $x=1$ is the asymptotic infinity,
while $y=-1$ gives
the south pole semi-axis and 
$y=1$ the north pole semi-axis.

Next we parametrize the metric perturbation
functions so that we can restrict to solutions that are outgoing waves at infinity and ingoing waves at the horizon. An appropriate choice for the metric perturbations is
\begin{eqnarray}
 H_1 &=& \widetilde H_1(x,y)  \, \frac{1}{x(1-x)}  \, (1-y^2)^{M_z/2}  \, e^{i \hat{R}} \, ,
 \label{H1}
 \\
 T &=& \widetilde T(x,y)  \, (1-y^2)^{M_z/2} \,  \,  e^{i \hat{R}} \, , \\
 N &=& \widetilde N(x,y)  \, \frac{1}{x(1-x)}  \, (1-y^2)^{M_z/2}  \, e^{i \hat{R}} \, , \\
 L &=& \widetilde L(x,y) \,  \frac{1}{x(1-x)}  \, (1-y^2)^{M_z/2}  \, e^{i \hat{R}} \, , \\
 h_0 &=& \widetilde h_0(x,y) \,  \frac{1}{1-x} \,  (1-y^2)^{M_z/2}  \, e^{i \hat{R}} \, , \\
 h_1 &=& \widetilde h_1(x,y) \,  \frac{1}{x(1-x)} \,  (1-y^2)^{M_z/2}  \, e^{i \hat{R}} \, .
   \label{metric_param}
\end{eqnarray}
For the scalar perturbation we choose
\begin{eqnarray}
\Phi = \widetilde \Phi_1(x,y)  \, (1-x) \,  (1-y^2)^{M_z/2} \,  e^{i \hat{R}} \, .
\end{eqnarray}
The function $\hat{R}$ is chosen such that the perturbations satisfy the outgoing wave conditions at $x=1$ and the ingoing wave conditions at $x=0$.

The metric perturbation equations to be solved are in the components of the linearized Einstein equation, $\delta\mathcal{G}_{\mu\nu}=0$, which has 10 components. 
There are 6 undetermined perturbation functions for the metric, $\{\widetilde H_1, \widetilde T, \widetilde N, \widetilde L, \widetilde h_0, \widetilde h_1\}$. Additionally there is the scalar perturbation $\widetilde \Phi_1$, which is decoupled from the other perturbations since we are considering a scalar test field.
We proceed by simply choosing 
the following 6 components of the Einstein equation for numerical integration:
$\{ \delta\mathcal{G}_{tr}, \delta\mathcal{G}_{t\theta},  \delta\mathcal{G}_{rr},  \delta\mathcal{G}_{r\theta},  \delta\mathcal{G}_{r\phi},  \delta\mathcal{G}_{\theta\phi} \}$. 
The scalar equation is given by $\delta\mathcal{S}$, equation (\ref{scalar_eq}).

{
It is possible to choose other combinations of components for the numerical integration, however, our numerical calculations indicate that the results do not change significantly. 
In fact,
an inspection of the equations reveals that the above combination of components is slightly simpler than other combinations. For instance, %
in the static case, 
the above system of equations reduces to a system of ordinary differential equations (ODEs). It is formed of 2 first order ODEs for the axial functions $\widetilde h_0$ and $\widetilde h_1$, 2 first order ODEs for the polar functions $\widetilde T$ and $\widetilde H_1$, and 2 algebraic equations for $\widetilde N$ and $\widetilde L$,
while
the remaining Einstein tensor components include second order derivatives of the perturbation functions. In the presence of rotation, we have a complicated system of PDEs, but this choice of components contains lower order of derivatives {
as compared to other choices.}

}

Defining a  vector consisting of the perturbation functions, 
$\vec{X}=[\widetilde H_1, \widetilde T, \widetilde N, \widetilde L, \widetilde h_0, \widetilde h_1, \widetilde \Phi_1]$, 
the system of 7 linear and homogeneous PDEs in 
the compactified 
coordinates $(x,y)$ can be written as
\begin{eqnarray}
    \mathcal{D}_{\mathrm{I}}(x,y) \vec{X}(x,y) = 0, \, \, \, \, \quad   \mathrm{I} = 1,...,7 \, .
    \label{metric_eq_xy}
\end{eqnarray}
Equation (\ref{metric_eq_xy}) must be satisfied in the bulk of the domain,
while on the 4 boundaries of the domain we need to impose appropriate conditions, which are discussed in the next subsection.

\subsection{Boundary conditions}

{
Next we require that the perturbation functions behave like an outgoing wave at infinity. This means that the perturbation functions must have the well-known form of
\begin{eqnarray}
           T &=& e^{i\omega R^* } \left( T^{+}(\theta) + \mathcal{O}\left(\frac{1}{r}\right) \right) \, ,\\
           H_{1} &=& r e^{i\omega R^* } \left( H^{+}_{1}(\theta) + \mathcal{O}\left(\frac{1}{r}\right) \right)\, ,\\
           N &=& r e^{i\omega R^* } \left( N^{+}(\theta) + \mathcal{O}\left(\frac{1}{r}\right)  \right) \, ,\\
           L &=& r e^{i\omega R^* } \left( L^{+}(\theta) + \mathcal{O}\left(\frac{1}{r}\right)  \right) \, ,\\
           h_{0} &=&  r e^{i\omega R^* } \left( h^{+}_{0}(\theta) + \mathcal{O}\left(\frac{1}{r}\right) \right)\, ,\\
           h_{1} &=&  r e^{i\omega R^* } \left( h^{+}_{1}(\theta)  + \mathcal{O}\left(\frac{1}{r}\right) \right)\, ,\\
           \Phi_{1} &=& \frac{1}{r} e^{i\omega R^* } \left(  \Phi_{1}^{+}(\theta) + \mathcal{O}\left(\frac{1}{r}\right)   \right)\, ,
\end{eqnarray}
where we have expressed them in $(r, \theta)$ variables for clarity, and $ \frac{dR^*}{dr} =  1 + \frac{2M}{r} + \mathcal{O}\left(\frac{1}{r^2}\right)  $.

Introducing these expansions into the perturbation equations, 
}
we obtain a set of 6 boundary conditions for the metric perturbations, plus a condition for the scalar. 
{
Using the parametrization from the previous subsection,}
some of these conditions are particularly simple, such as,
\begin{eqnarray}
i\widetilde L|_{x=1} + r_H \omega \widetilde T|_{x=1} = 0 \, , \\
r_H \omega \widetilde T|_{x=1} - i \widetilde N|_{x=1} - 2 i \widetilde H_1|_{x=1} = 0 \, .
\end{eqnarray}
These are the same structural conditions one finds in the static case.
Other relations are, however, more complicated, e.g.~involving derivatives in $r$ and $\theta$ of the functions. 
In general the conditions can be written in an operator form,
\begin{eqnarray}
    \mathcal{A}_{\mathrm{I}}(x,y) \vec{X}(x,y)|_{x=1} = 0, \, \, \, 
    \quad 
    \mathrm{I} = 1,...,7 \, ,
\label{bcg_inf}
\end{eqnarray}
where $\mathcal{A}_{\mathrm{I}}$ are linear operators in $(x,y)$.

{
The procedure at the horizon is similar. We require the perturbation functions to be ingoing waves at the horizon, meaning
\begin{eqnarray}
           T &=& e^{-i(\omega-M_z\Omega_H) R^* } \left( T^{-}(\theta)  + 
           \mathcal{O}\left(r-r_H\right) \right)\, ,\\
           H_{1} &=& \frac{r_H}{r-r_H} e^{-i(\omega-M_z\Omega_H) R^* }  \left( H^{-}_{1}(\theta) + \mathcal{O}\left(r-r_H\right) \right)\, ,\\
           N &=& \frac{r_H}{r-r_H} e^{-i(\omega-M_z\Omega_H) R^* } \left( N^{-}(\theta)  + \mathcal{O}\left(r-r_H\right) \right)\, ,\\
           L &=& \frac{r_H}{r-r_H} e^{-i(\omega-M_z\Omega_H) R^* }  \left( L^-(\theta) + \mathcal{O}\left(r-r_H\right) \right)\, ,\\
           h_{0} &=& e^{-i(\omega-M_z\Omega_H) R^* }  \left( h^{-}_{0}(\theta) + \mathcal{O}\left(r-r_H\right) \right)\, ,\\
           h_{1} &=& \frac{r_H}{r-r_H} e^{-i(\omega-M_z\Omega_H) R^* }  \left( h^{-}_{1}(\theta) + \mathcal{O}\left(r-r_H\right) \right)\, ,\\
           \Phi_{1} &=& \frac{1}{r} e^{-i(\omega-M_z\Omega_H) R^* }  \left(  \Phi_{1}^-(\theta) + \mathcal{O}\left(r-r_H\right) \right)\, , 
\end{eqnarray}
where
$\frac{dR^*}{dr} = \frac{r_H(a^2 + r_H^2)}{(r_H^2-a^2)(r-r_H)} + \mathcal{O}(1) $.

This dictates that
}
an ingoing wave solution must satisfy relations that can be written in a similar form by defining another set of linear operators $\mathcal{B}_{\mathrm{I}}$,
\begin{eqnarray}
    \mathcal{B}_{\mathrm{I}}(x,y) \vec{X}(x,y)|_{x=0} = 0, \, \, \,   \quad \mathrm{I} = 1,...,7 \, .
\label{bcg_hor}
\end{eqnarray}

In addition, we have to ensure regularity at the north pole and south pole semi-axis. 
{
We require the perturbation functions to have the following regular expansion at $y=1$, the north pole semi-axis,
\begin{eqnarray}
           T &=& T^{NP}(x)  + \mathcal{O}\left(y-1\right)   \, ,\\
           H_{1} &=& H_1^{NP}(x)  + \mathcal{O}\left(y-1\right)   \, ,\\
           N &=& N^{NP}(x)  + \mathcal{O}\left(y-1\right)   \, ,\\
           L &=& L^{NP}(x)  + \mathcal{O}\left(y-1\right)   \, ,\\
           h_0 &=& h_0^{NP}(x)  + \mathcal{O}\left(y-1\right)   \, ,\\
           h_1 &=&  h_1^{NP}(x)  + \mathcal{O}\left(y-1\right)   \, ,\\
           \Phi_{1} &=& \Phi_{1}^{NP}(x)  + \mathcal{O}\left(y-1\right)   \, ,
\end{eqnarray}
}
and we find another set of relations that can be written in an operator form
\begin{eqnarray}
    \mathbf{\alpha}_{\mathrm{I}}(x,y) \vec{X}(x,y)|_{y=1} = 0, \, \, \, 
     \quad \mathrm{I} = 1,...,7 \, .
\label{bcg_np}
\end{eqnarray}
{
Similarly, we require the perturbation functions to be regular at $y=-1$, the south pole semi-axis,
\begin{eqnarray}
           T &=& T^{SP}(x)  + \mathcal{O}\left(y+1\right)   \, ,\\
           H_{1} &=& H_1^{SP}(x)  + \mathcal{O}\left(y+1\right)   \, ,\\
           N &=& N^{SP}(x)  + \mathcal{O}\left(y+1\right)   \, ,\\
           L &=& L^{SP}(x)  + \mathcal{O}\left(y+1\right)   \, ,\\
           h_0 &=& h_0^{SP}(x)  + \mathcal{O}\left(y+1\right)   \, ,\\
           h_1 &=&  h_1^{SP}(x)  + \mathcal{O}\left(y+1\right)   \, ,\\
           \Phi_{1} &=& \Phi_{1}^{SP}(x)  + \mathcal{O}\left(y+1\right)   \, ,
\end{eqnarray}
and find another set of relations of the form
}
\begin{eqnarray}
    \mathbf{\beta}_{\mathrm{I}}(x,y) \vec{X}(x,y)|_{y=-1} = 0, \, \, \, 
     \quad \mathrm{I} = 1,...,7 \, .
\label{bcg_sp}
\end{eqnarray}

\subsection{Spectral decomposition}

At this stage we have the system of PDEs and the boundary conditions that describe the quasinormal mode perturbations. In order to solve the problem numerically, we use a spectral method.
Thus we decompose the metric perturbations in a series of Chebyshev polynomials and Legendre functions, %
\begin{eqnarray}
    \widetilde H_1(x,y) &=& \sum_{k=0}^{N_x-1}  \, \, \sum_{l=|M_z|}^{N_y+|M_z|-1} C_{1,k,l}  \,  T_k(x)  \,  P_l^{M_z}(y)  \,  (1-y^2)^{-M_z/2} \, , \\
    \widetilde T(x,y) &=& \sum_{k=0}^{N_x-1}  \, \, \sum_{l=|M_z|}^{N_y+|M_z|-1} C_{2,k,l}  \,  T_k(x)  \, 
 P_l^{M_z}(y)  \,  (1-y^2)^{-M_z/2} \, , \\
    \widetilde L(x,y) &=& \sum_{k=0}^{N_x-1}  \, \, \sum_{l=|M_z|}^{N_y+|M_z|-1} C_{3,k,l}  \,  T_k(x)  \,  P_l^{M_z}(y)  \,  (1-y^2)^{-M_z/2} \, , \\
    \widetilde N(x,y) &=& \sum_{k=0}^{N_x-1}  \, \, \sum_{l=|M_z|}^{N_y+|M_z|-1} C_{4,k,l}  \,  T_k(x)  \,  P_l^{M_z}(y)  \,  (1-y^2)^{-M_z/2} \, , \\
    \widetilde h_0(x,y) &=& \sum_{k=0}^{N_x-1}  \, \, \sum_{l=|M_z|}^{N_y+|M_z|-1} C_{5,k,l}  \,  T_k(x)  \,  P_l^{M_z}(y)  \,  (1-y^2)^{-M_z/2} \, , \\
    \widetilde h_1(x,y) &=& \sum_{k=0}^{N_x-1}  \, \, \sum_{l=|M_z|}^{N_y+|M_z|-1} C_{6,k,l}  \,  T_k(x)  \, 
 P_l^{M_z}(y)  \,  (1-y^2)^{-M_z/2} \, .
    \label{metric_dec}
\end{eqnarray}
We proceed in the same way for the scalar perturbation,
\begin{eqnarray}
    \widetilde \Phi_1(x,y) &=& \sum_{k=0}^{N_x-1} \, \, \sum_{l=|M_z|}^{N_y+|M_z|-1}  C_{7,k,l}  \,  T_k(x)  \,  P_l^{M_z}(y)  \,  (1-y^2)^{-M_z/2} \, .
\end{eqnarray}
We have decomposed the radial behaviour of the perturbation functions
in $T_k(x)$ functions, which are the
Chebyshev polynomials of the first kind. 
The angular dependence is decomposed in the $P_l^{M_z}$ functions, which are the Legendre functions of the first kind. 

The constants $C_{n,k,l}$ are to be determined by solving the PDEs subject to the boundary conditions. Note that there are a total of $6 \times N_x \times N_y$ undetermined constants for the metric perturbations, plus an additional $N_x \times N_y$ constants for the scalar perturbation.

The next step of the spectral method is to discretize the domain of integration. 
For the $x$-coordinates we choose the Gauss-Lobatto points,
\begin{eqnarray}
x_I = \frac{1}{2} \left( 1+\cos{\left(\frac{I-1}{N_x-1}\pi\right)} \right) \, , \, \quad  I=1,...,N_x \, . 
\end{eqnarray}
As for the $y$-coordinates, we choose uniformly separated points, 
\begin{eqnarray}
   y_K = 2\frac{K-1}{N_y-1}-1 
   \, , \, \quad K=1,...,N_y \, . 
\end{eqnarray}
This forms a grid of $N_x \times N_y$ points. The next step is to evaluate the equations at every point of this grid. 

{
In principle it is possible to use other distributions of points for the grids. For example, we have also tested uniformly separated points for both coordinates, obtaining very similar results. However, it is well known that typically the choice of the Gauss-Lobato points optimizes the numerical calculations when dealing with Chebyshev polynomials, and we have observed in some cases a significant reduction in the calculation time 
when using the Gauss-Lobato points as compared to the uniformly distributed points.
}

On the boundaries,
we evaluate the corresponding boundary conditions, 
while in the bulk of the domain we evaluate the PDEs.
Note that at each point of the grid we evaluate the 6 metric perturbation equations plus the scalar equation. 
This means that, after evaluating the equations at each point of the grid, we have a total of $7 \times N_x \times N_y$ algebraic equations. These equations can be written in a matrix form in the following way,
\begin{eqnarray}
    \left( \mathcal{M}_0 + \mathcal{M}_1 \omega + \mathcal{M}_2 \omega^2 \right) \Vec{C} = 0 \, .
    \label{matrix_eq}
\end{eqnarray}
Here $\vec{C}$ is a vector 
consisting of all the constants
$C_{a,k,l}$ with $a=1,...,7$, and $\mathcal{M}_0$, $\mathcal{M}_1$ and $\mathcal{M}_2$ are square matrices of size $(7 \times N_x \times N_y) \times (7 \times N_x \times N_y)$.

Equation (\ref{matrix_eq}) has the form of a standard quadratic eigenvalue problem, where the eigenvalue is $\omega$. In order to obtain the quasinormal modes and the corresponding eigenvectors $\vec C$, we have implemented the numerical procedures in both Maple and Matlab with the Multiprecision Computing Toolbox Advanpix \cite{Advanpix}.

To numerically cross-check our results, we evaluate the resulting quasinormal modes and the corresponding perturbation functions in the remaining set of PDEs that we did not use for the spectral decomposition. They are the 4 equations $\{ \delta\mathcal{G}_{tt}, \delta\mathcal{G}_{t\phi},  \delta\mathcal{G}_{\theta\theta},  \delta\mathcal{G}_{\phi\phi} \}$, that are required to be satisfied at each point of the grid with a tolerance 
{
smaller than $10^{-4}$}.
{
In this way we guarantee that the resulting quasinormal modes are physical.
}

As shown in the next section, this method not only allows for the calculation of the fundamental modes, but it also computes several excited modes with good precision. In addition, it also allows us to study modes of different leading multipoles. 

\section{Results}
\label{results}

\subsection{Scalar modes}

\begin{figure}
    \centering
    
    \subfloat[]{\includegraphics[angle=-90,width=0.45\textwidth]{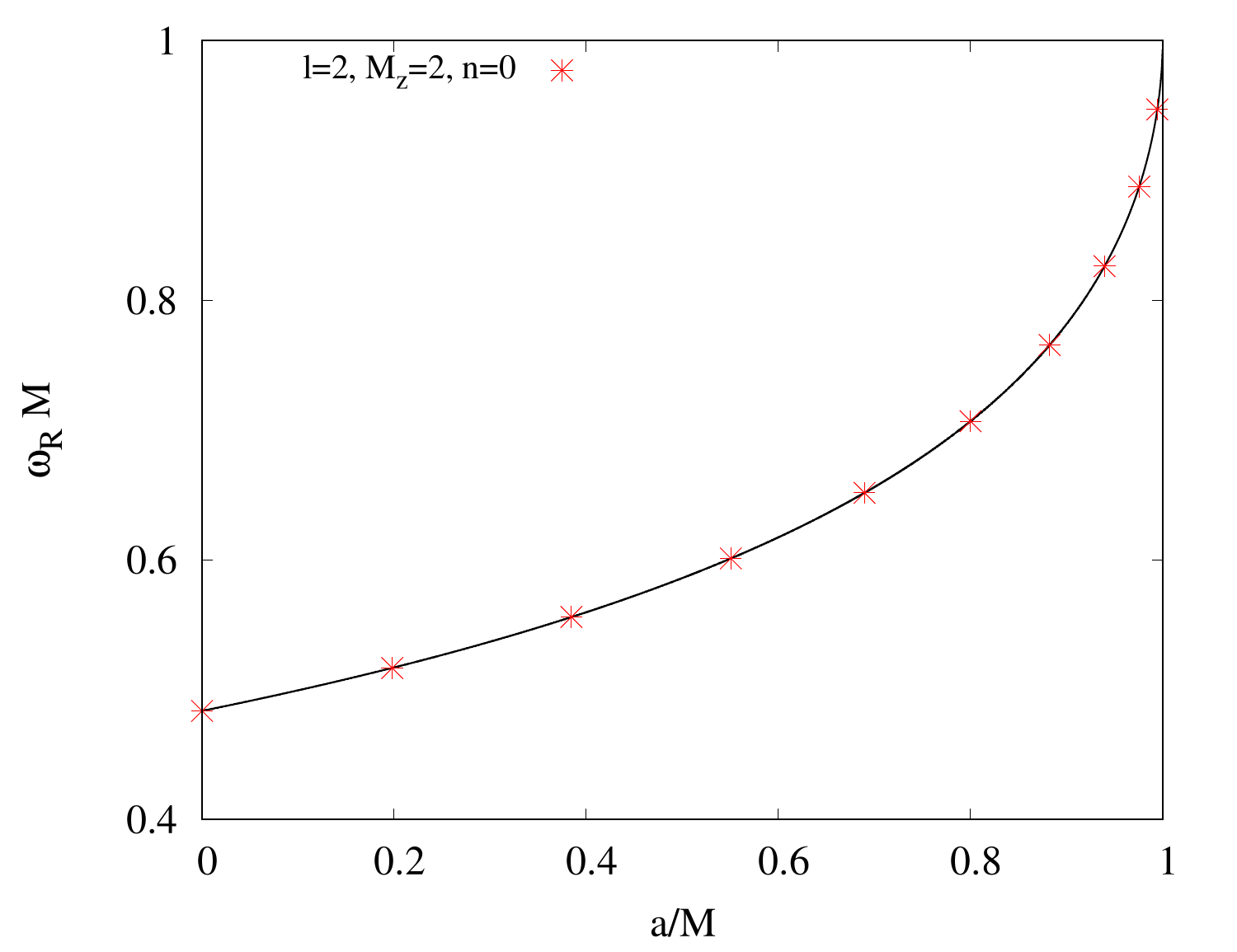}
    }
    \subfloat[]{\includegraphics[angle=-90,width=0.45\textwidth]{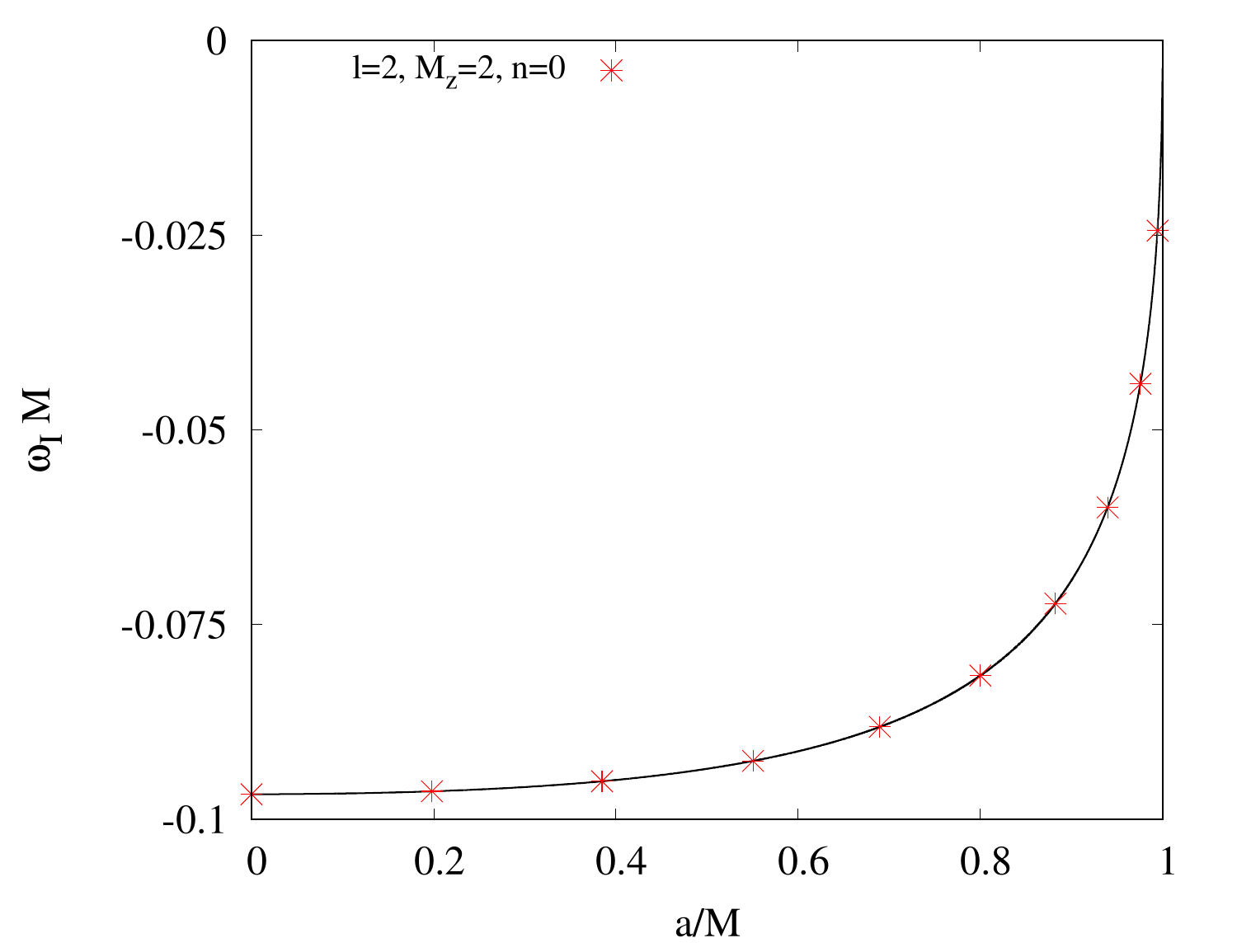}
    }
    \caption{Scalar $l=2, M_z=2$ fundamental modes: (a) Real frequency part scaled with the mass $\omega_R M$ versus the Kerr parameter inversely scaled with the mass $a/M$. In red are results from the spectral method with $N_x=18$ and $N_y=14$, and the solid line constitutes the well-known results from the Teukolsky equation.
    (b) Analogous plot for the imaginary frequency part scaled with the mass $\omega_I M$.}
    \label{scalar_l2m2n0}
\end{figure}

We present first the results for the scalar perturbation, which can be treated as an independent problem as it is decoupled from the metric perturbations. 
Let us focus the discussion on the modes with $M_z=2$, and, in particular, on the fundamental $(l=2)$-led mode. 
By $(l=2)$-led mode, we mean the mode that connects to the purely $l=2$ mode in the static limit, where spherical symmetry is restored.
In the following discussions, 
we will refer to the mode simply as the fundamental $l=2, M_z=2$ mode, but be aware that in general, when the background is spinning, the perturbation function $\Phi$ is a sum of different $l$-multipoles.  

Shown in Figure~\ref{scalar_l2m2n0} is
the dependence of the real $\omega_R$ and imaginary $\omega_I$ frequency parts on %
the Kerr parameter $a$, scaled with the black hole mass $M$. In particular, these results are obtained using a relatively small grid, with just $N_x=18$ and $N_y=14$. 
The red points are obtained by the spectral method, and the solid black line is the well--known result obtained from the Teukolsky equation \cite{Berti:2009kk}.

A comparison between both calculations reveals an excellent precision of the spectral method,
ranging from a relative error of the order of $10^{-8}$ for solutions with moderate angular momentum ($a/M<0.5$), to a $1\%$ error as we {almost} reach extremality at $a/M=0.995$. 
These results can be easily improved by increasing the number of grid points as we approach extremality.

\subsection{Metric modes}

Here we present 
the core results for the full metric perturbations of Kerr
using the spectral method.

\begin{figure}
    \centering
    \includegraphics[angle=-90,width=0.5\textwidth]{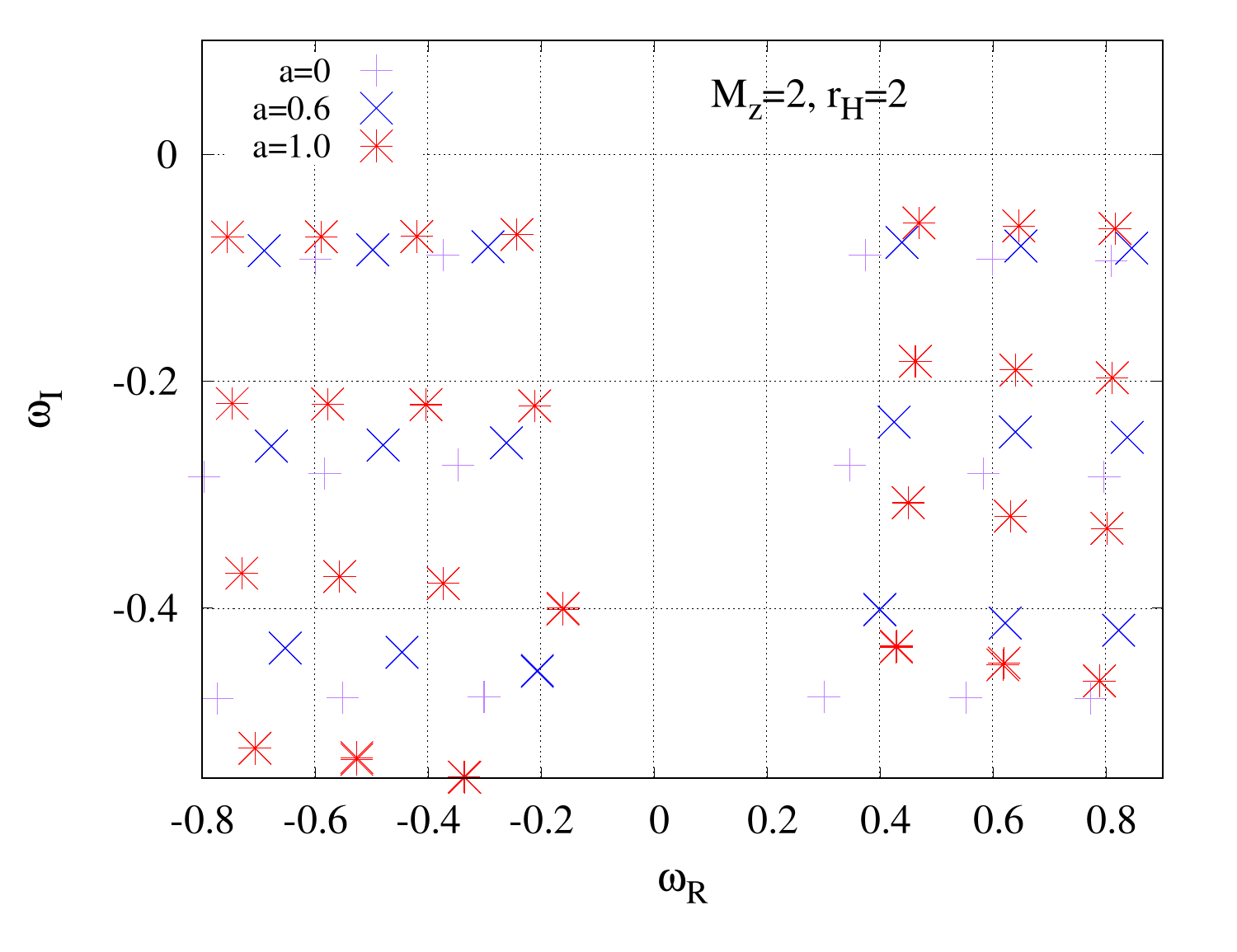}
    \caption{An example of a quasinormal mode spectrum for Kerr black holes.}
    \label{fig:spectrum_example}
\end{figure}

In Figure \ref{fig:spectrum_example}
we show a section of the typical quasinormal mode spectrum that can be generated with our spectral method. In particular we show the 
$M_z=2$ spectrum for $r_H=2$, in the range of the real part of the frequency $-0.8 < \omega_R < 0.8$
and in the range of the imaginary part of the frequency
$\omega_I>-0.6$. We present the spectrum for 3 different values of the Kerr parameter $a$: 
{in purple we indicate with $+$ the $a=0$ (static) case,
in blue with $\times$ the $a=0.6$ case, and in red with $*$ (asterisks) the $a=1$ case. }
These results have been obtained for a grid of $N_x=N_y=20$.
Note that we can obtain modes with positive and negative real parts (co-rotating and counter-rotating modes respectively). We also obtain modes with different leading multipolar contribution and excitations. 
For this particular grid, the remaining set of PDEs is satisfied within $10^{-4}$ error or less at each point.

For example, let us focus on the modes for the $a=1$ black hole (with red asterisks). 
The first row of modes contains the fundamental modes $(n=0)$, namely the modes with the smallest value of $|\omega_I|$. 
The modes closest to the $\omega_R=0$ axis are those that are dominated by the quadrupolar spherical harmonics (i.e. $l=2$ modes in our terminology). 
As we move away from the $\omega_R=0$ axis, we get modes dominated by higher multipoles, i.e. $l=3,4,5$ modes, sequentially. 

Below the row of fundamental modes, we find other rows of modes, containing the excitations. 
In Figure \ref{fig:spectrum_example} we also show some of the $n=1,2,3$ and $n=4$ excitations. 
These excitations are ordered in a similar way as the fundamental modes: the modes closest to the $\omega_R=0$ axis are the $(l=2)$-led modes, and as we move away from the axis, we get modes dominated by higher multipoles, $l=3,4,...$\,.

The spectrum of modes for other values of the angular momentum exhibits a similar pattern, as is seen for the other values of $a$ shown in Figure
\ref{fig:spectrum_example}
(blue and purple points).

As the black hole spins faster, the absolute value of the imaginary part of the modes tends to decrease, i.e., the modes tend to be longer-lived.
In general there is a tendency for the modes to pile up as the angular momentum increases. 
As we will explain below, it becomes more challenging to extract the higher excitation modes and the higher multipoles as the spin increases towards extremality.

\begin{figure}[h]
    \centering
    
    \subfloat[]{\includegraphics[angle=-90,width=0.5\textwidth]{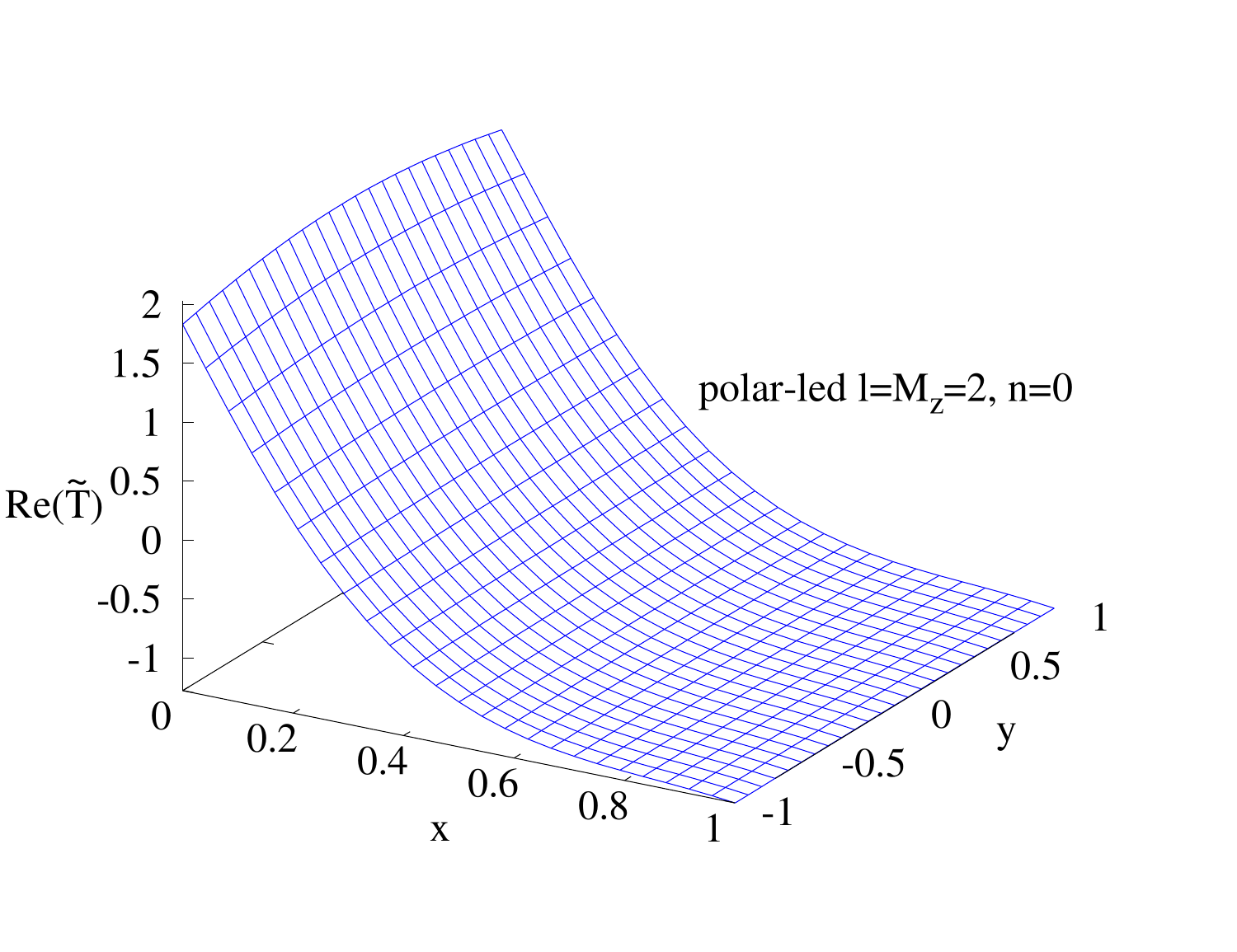}
    }
    \subfloat[]{\includegraphics[angle=-90,width=0.5\textwidth]{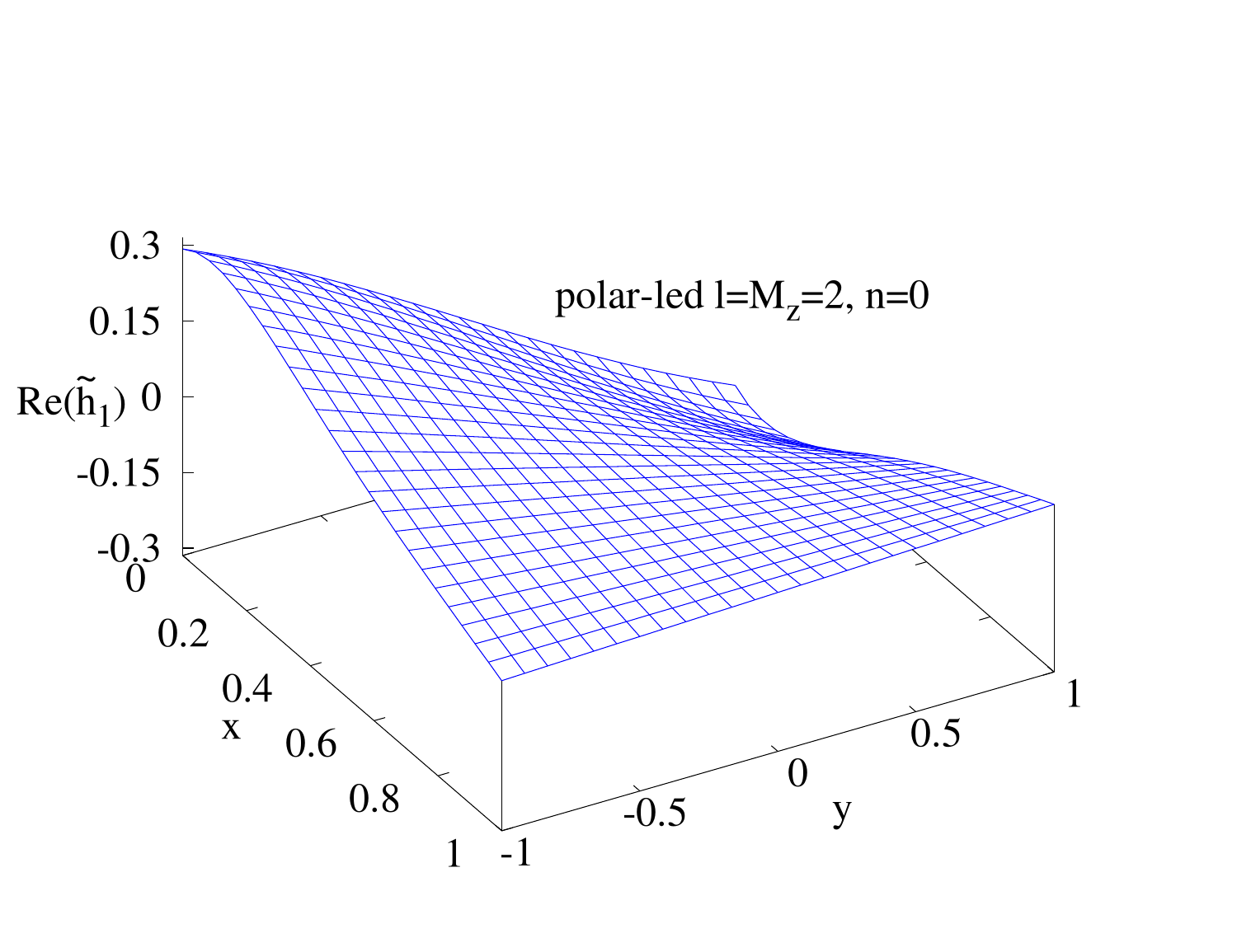}
    }
    \subfloat[]{\includegraphics[angle=-90,width=0.5\textwidth]{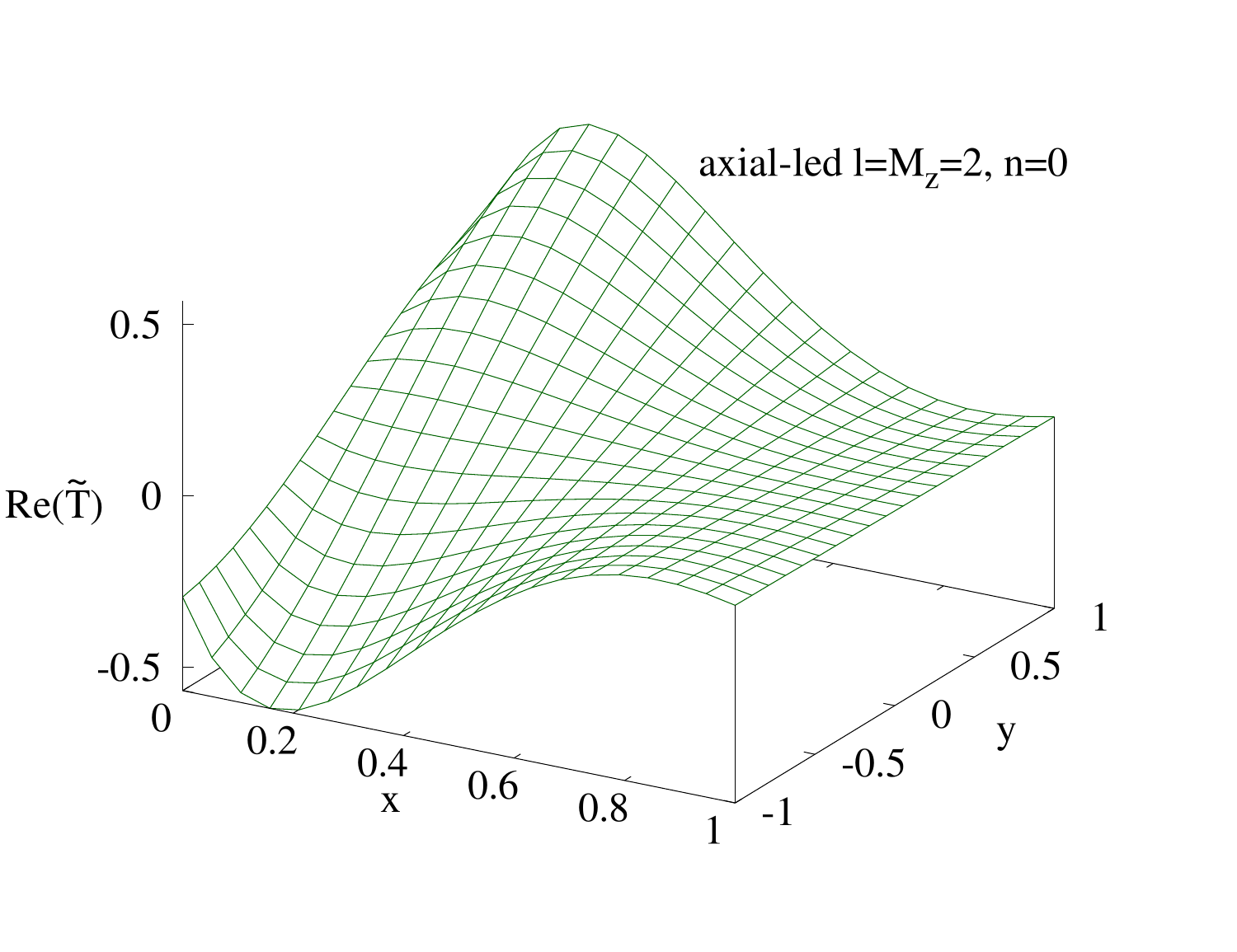}
    }
    \subfloat[]{\includegraphics[angle=-90,width=0.5\textwidth]{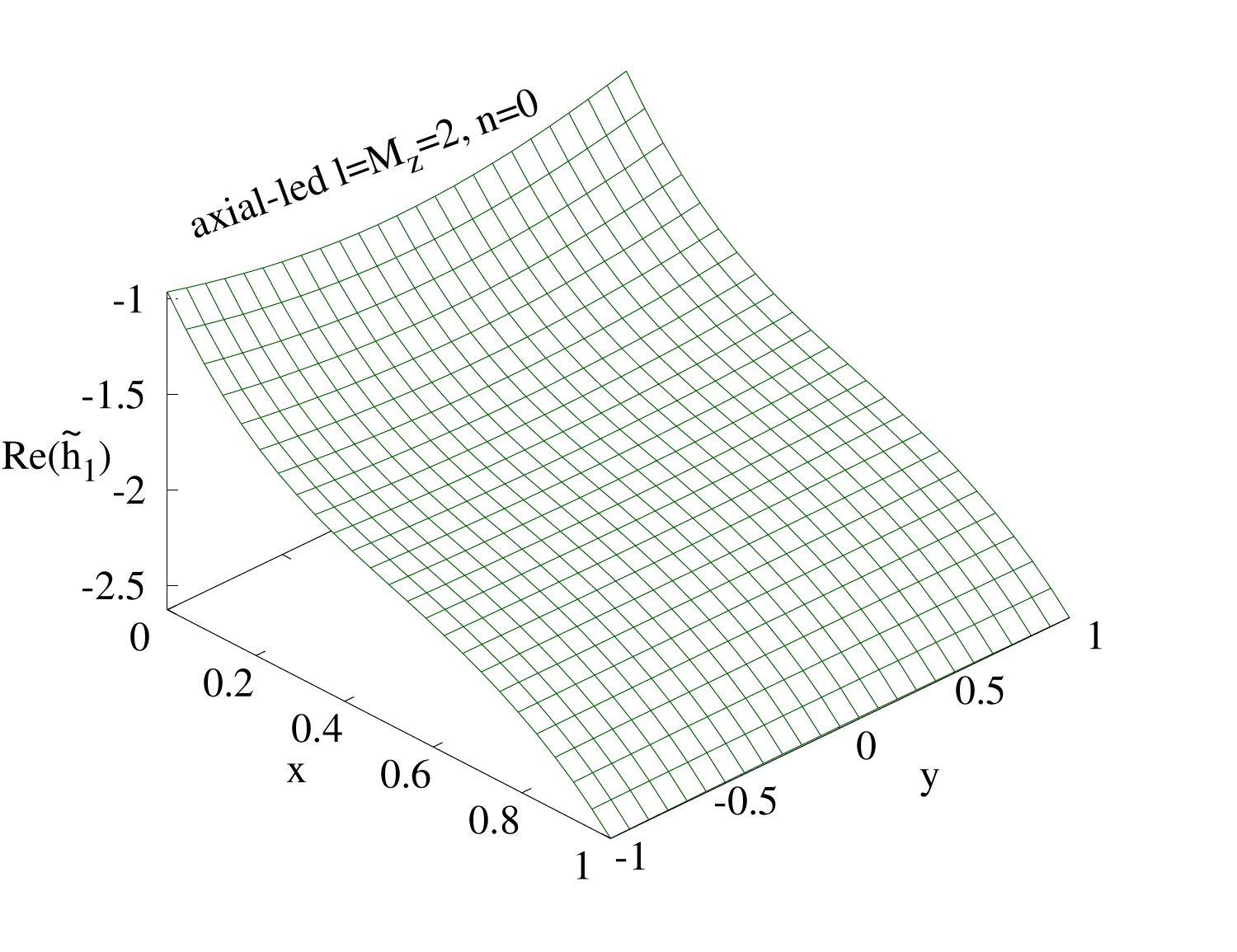}
    }
    \caption{Profile of the perturbation functions for the $l=M_z=2$ fundamental modes: (a) $\widetilde T$ function for the polar-led mode, (b) $\widetilde h_1$ function for the polar-led mode, (c) $\widetilde T$ function for the axial-led mode, (d) $\widetilde h_1$ function for the axial-led mode.}
    \label{profiles_all}
\end{figure}

It is important to note that there is a degeneracy in the modes that cannot be appreciated in Figure \ref{fig:spectrum_example}. 
Essentially, at each point we 
have two modes. 
One is the axial-led mode and the other is the polar-led mode. 
In order to distinguish if a mode is axial-led or polar-led, one has to look at the profile of the perturbation functions.

This brings us to 
Figures \ref{profiles_all}.
In Figures \ref{profiles_all}
we show the profiles of some representative perturbation functions. 
On the right panels we show the polar function $\widetilde T$ and on the left panels the axial function $\widetilde h_1$.
These perturbation functions correspond to the $l=M_z=2$ fundamental mode, for a configuration with $a=0.6$ and $r_H=2$.

Figures \ref{profiles_all} (a) and (b) show the profiles for the polar-led mode.
Since these figures correspond to the functions of the $(l=2)$-led mode,
the polar perturbation function $\widetilde T$
behaves predominantly like an even function with respect to the $y$-coordinate.
The rest of the polar perturbation functions behave similarly.
Meanwhile, the axial perturbation function
$\widetilde h_1$ behaves predominantly like an odd function
with respect to the $y$-coordinate. 
The other axial function behaves in the same manner as well.
This behaviour of the functions is exchanged in the axial-led perturbations, as it can be appreciated in Figures \ref{profiles_all} (c) and (d), showing again the $l=M_z=2$ fundamental mode for the same background solution. 
Note also that in the polar-led mode (top panels), the amplitude of the $\widetilde T$ function is significantly larger than that of the axial function $\widetilde h_1$, and vice versa for the axial-led mode (bottom panels). 
This is also useful in order to identify the nature of a mode.

\begin{figure}
    \centering
    \includegraphics[angle=-90,width=0.45\textwidth]{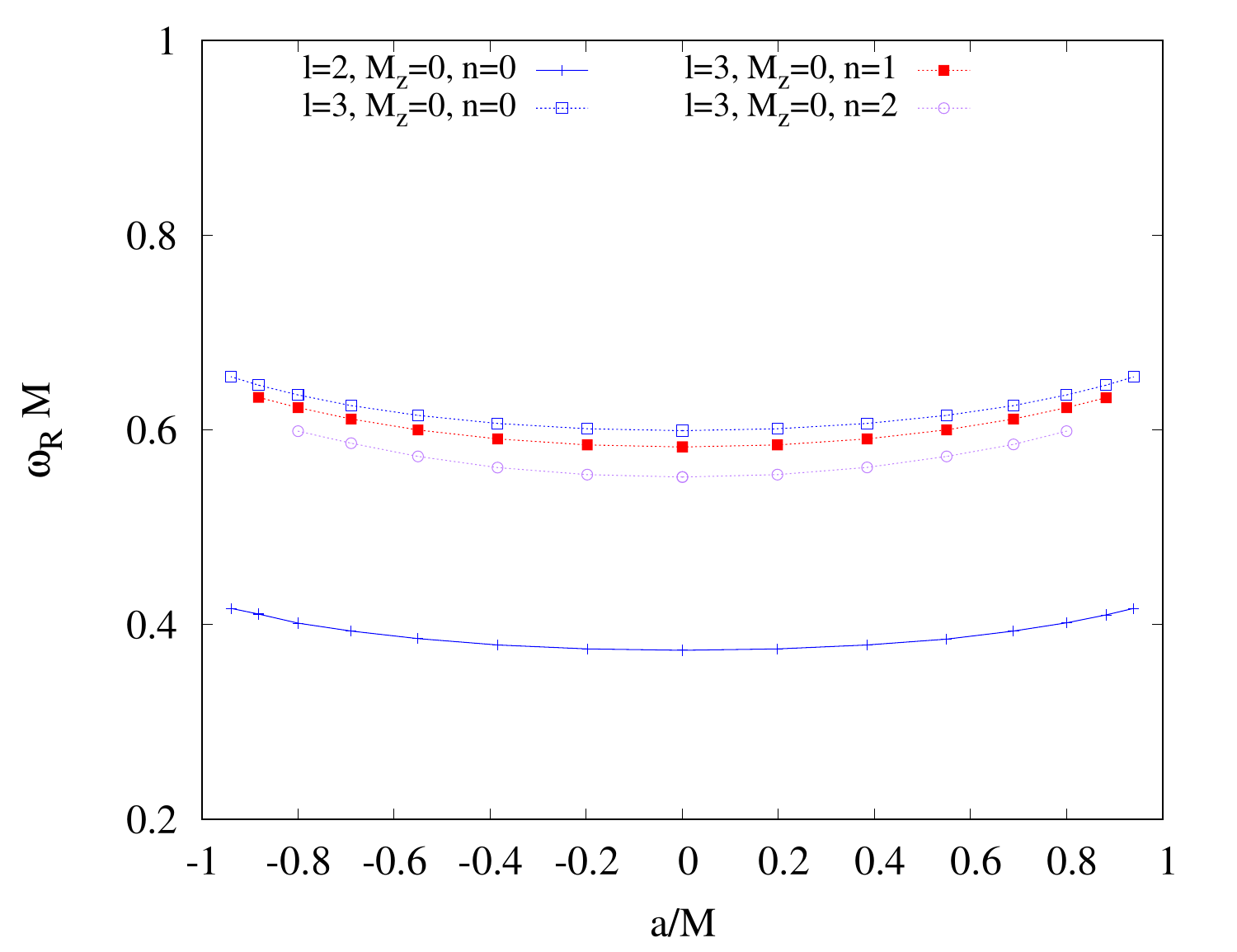}
    \includegraphics[angle=-90,width=0.45\textwidth]{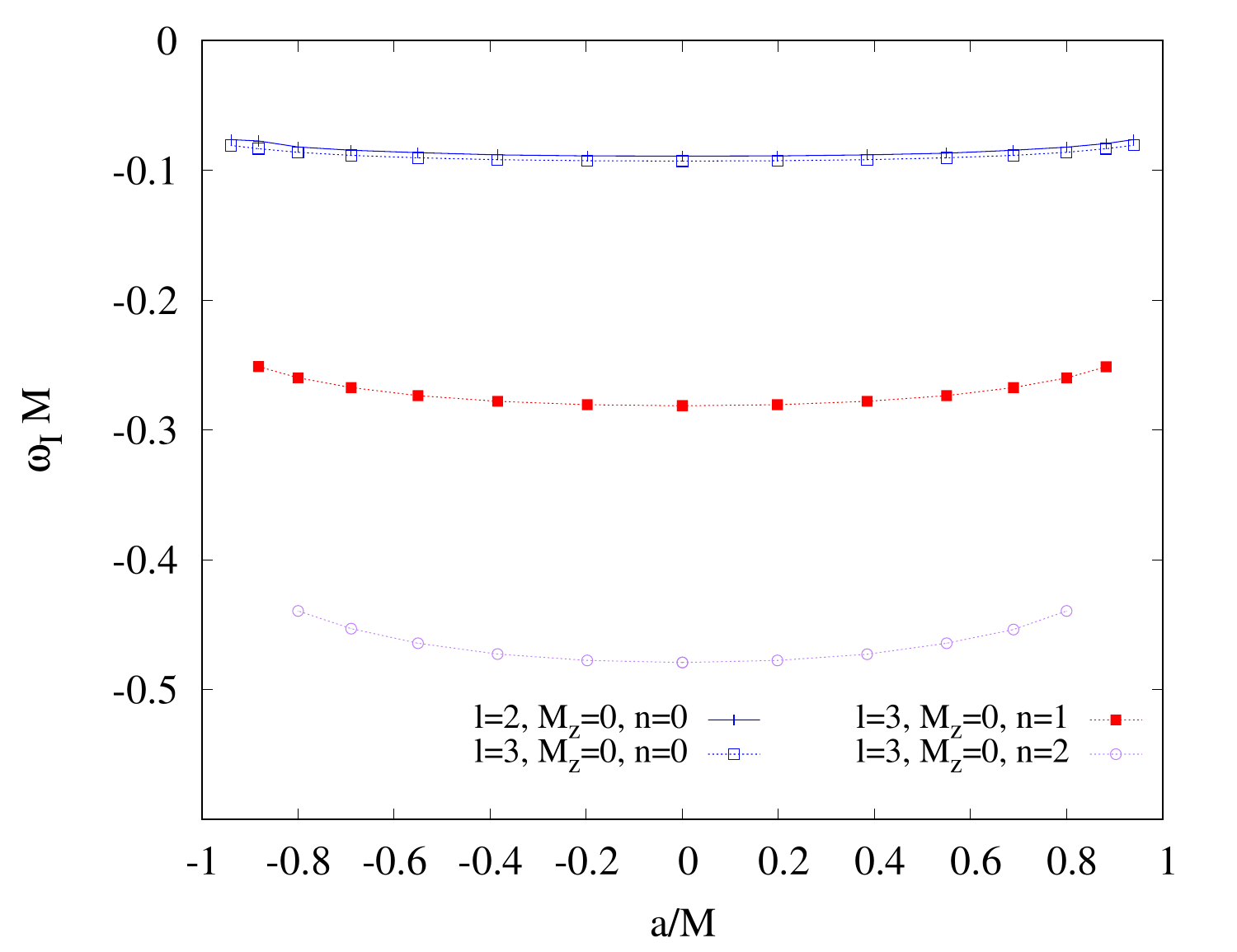}
    \includegraphics[angle=-90,width=0.45\textwidth]{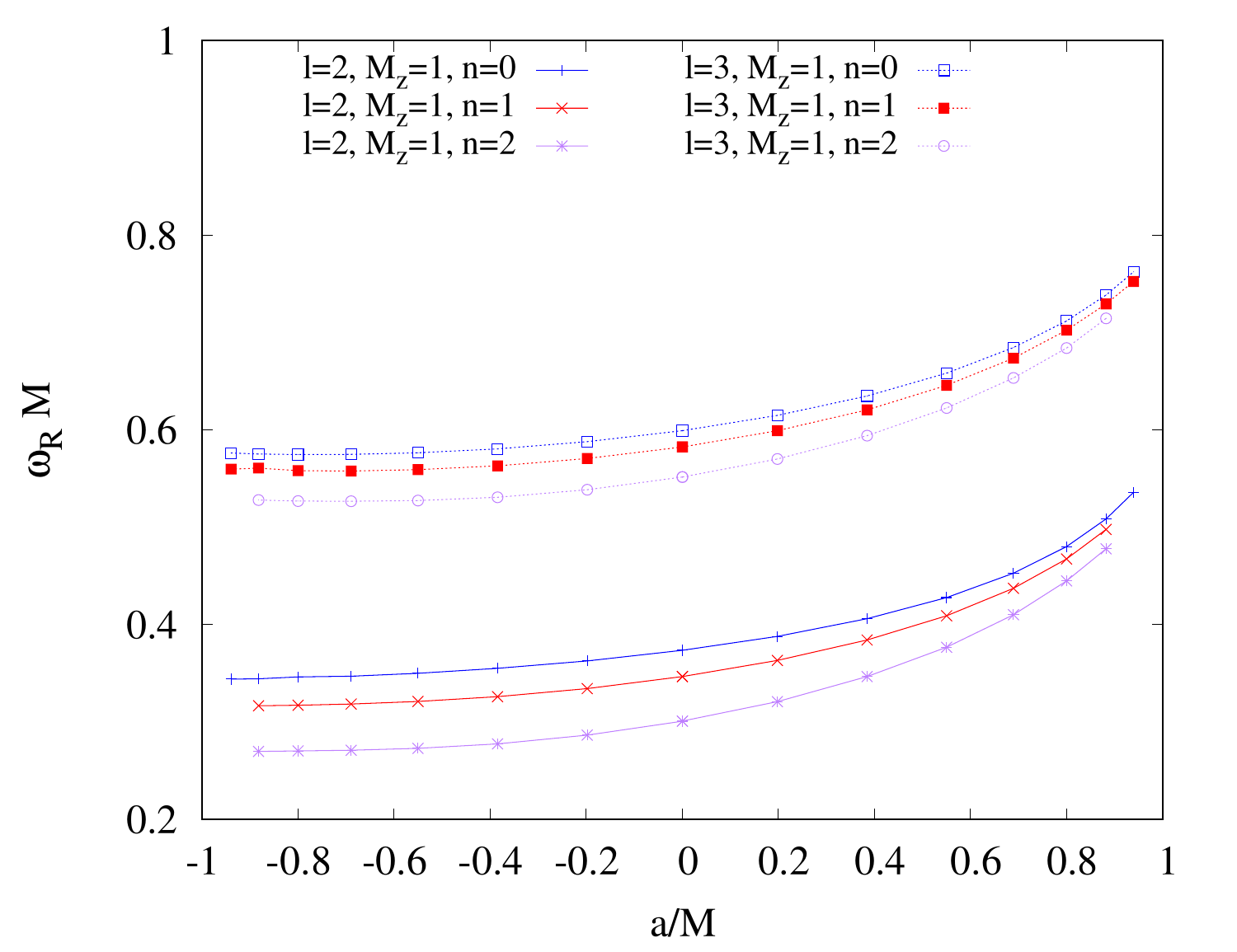}
    \includegraphics[angle=-90,width=0.45\textwidth]{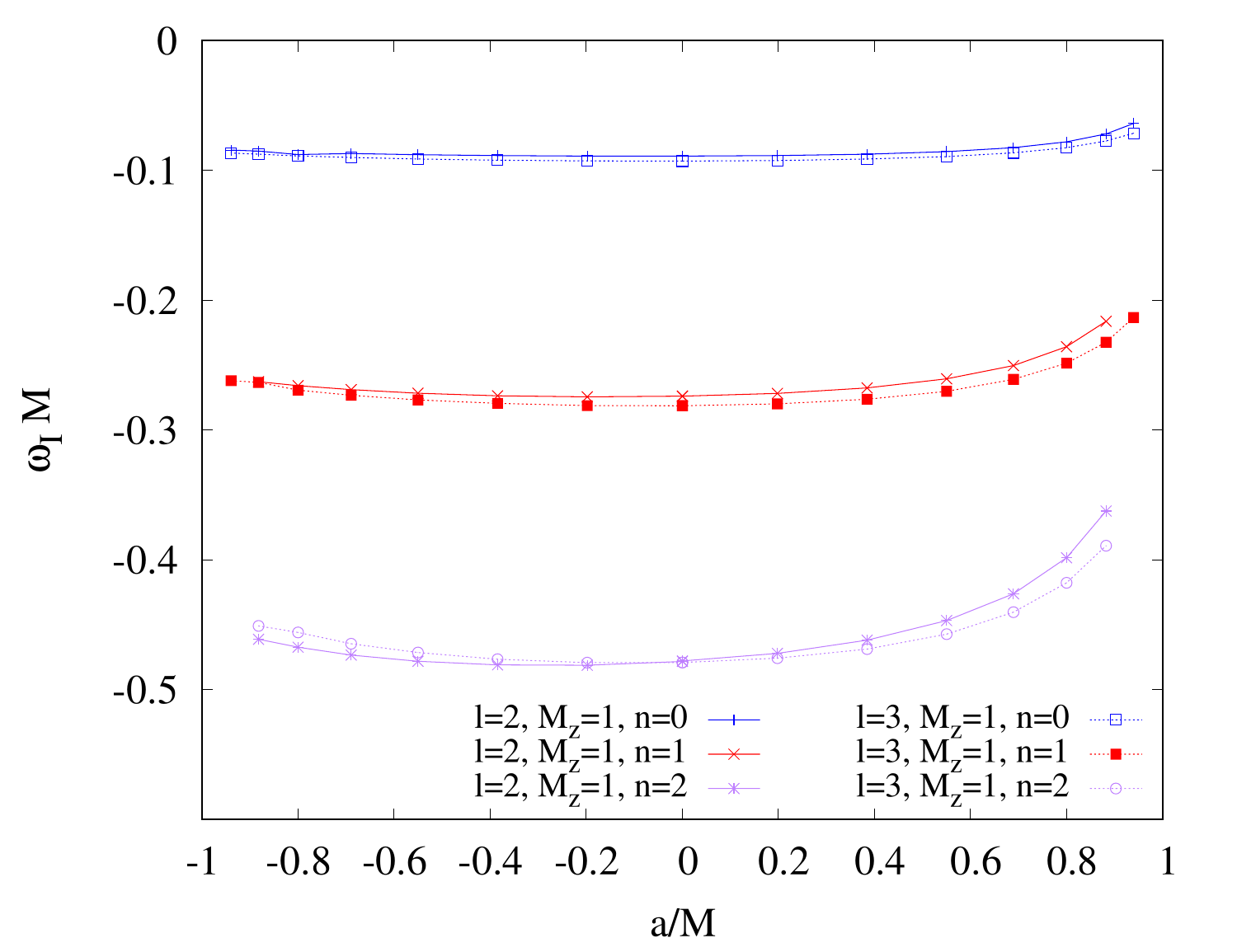}
    \includegraphics[angle=-90,width=0.45\textwidth]{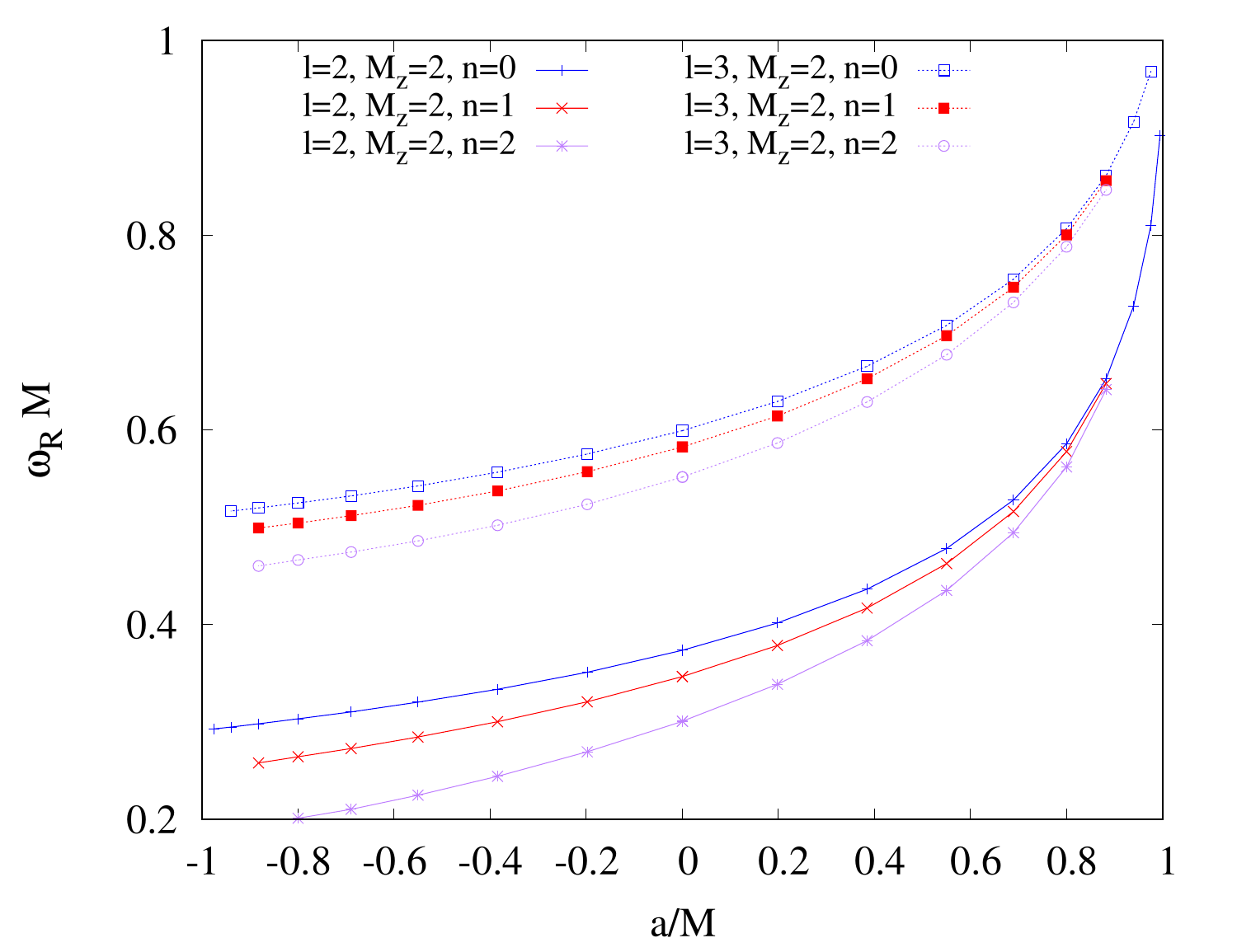}
    \includegraphics[angle=-90,width=0.45\textwidth]{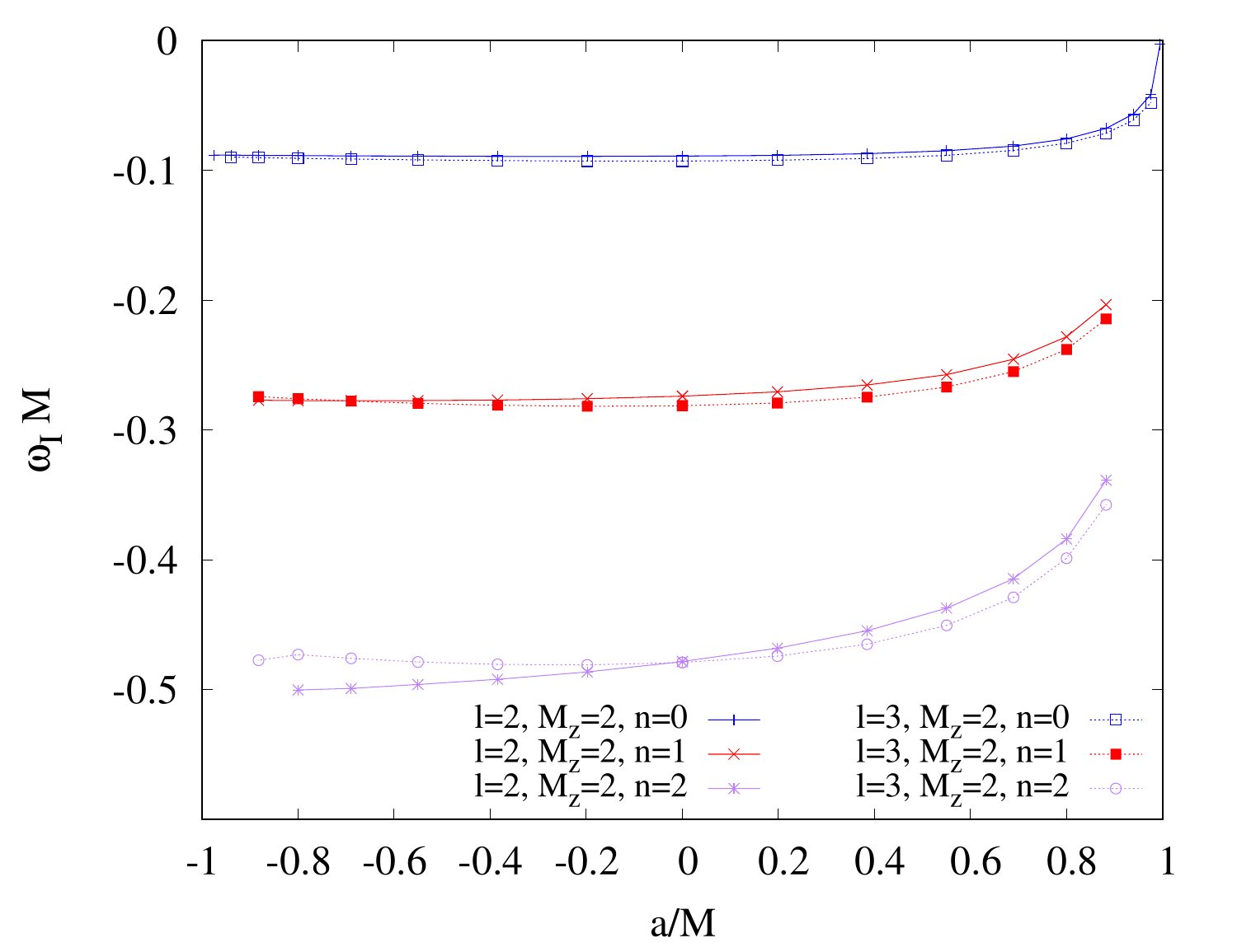}
    \includegraphics[angle=-90,width=0.45\textwidth]{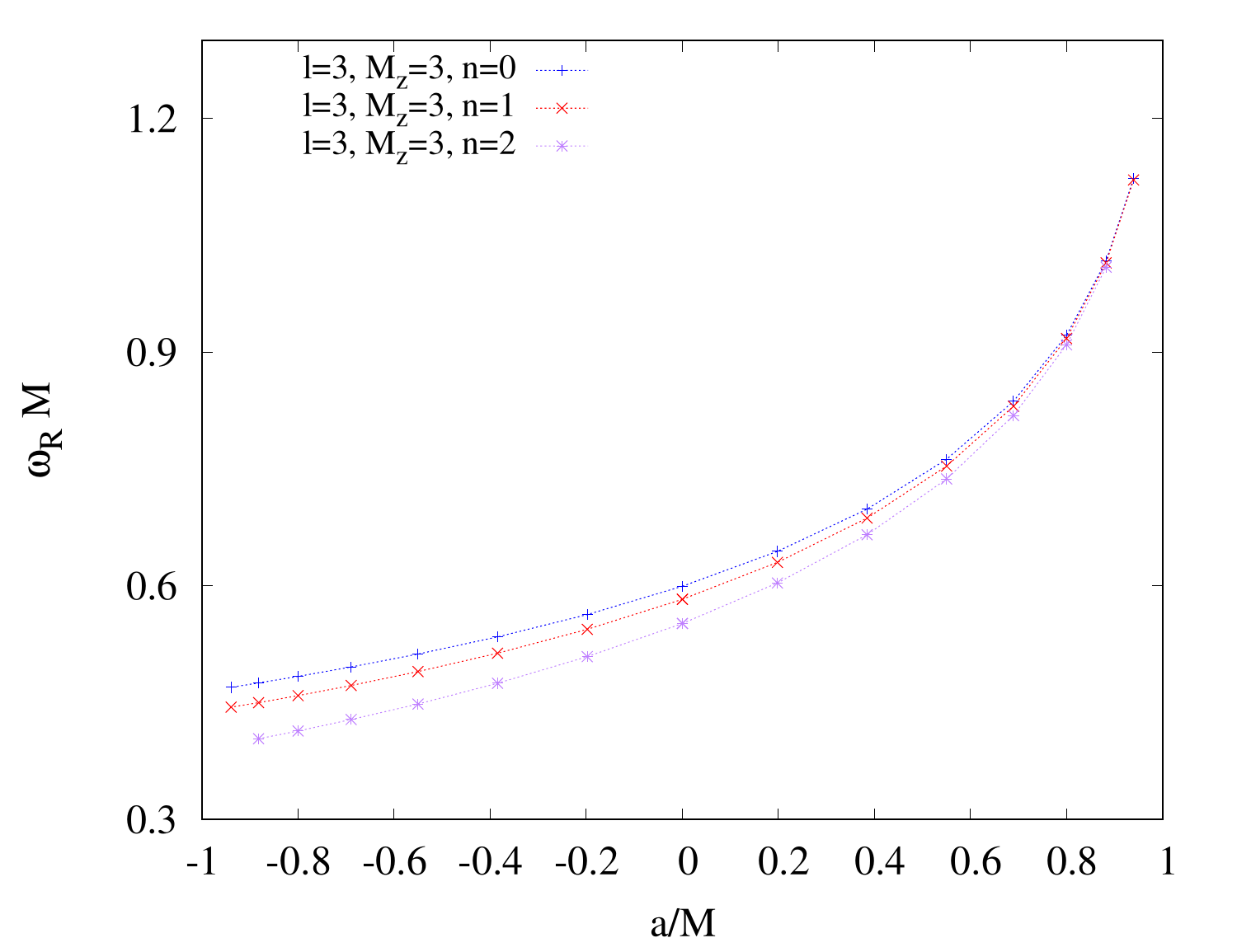}
    \includegraphics[angle=-90,width=0.45\textwidth]{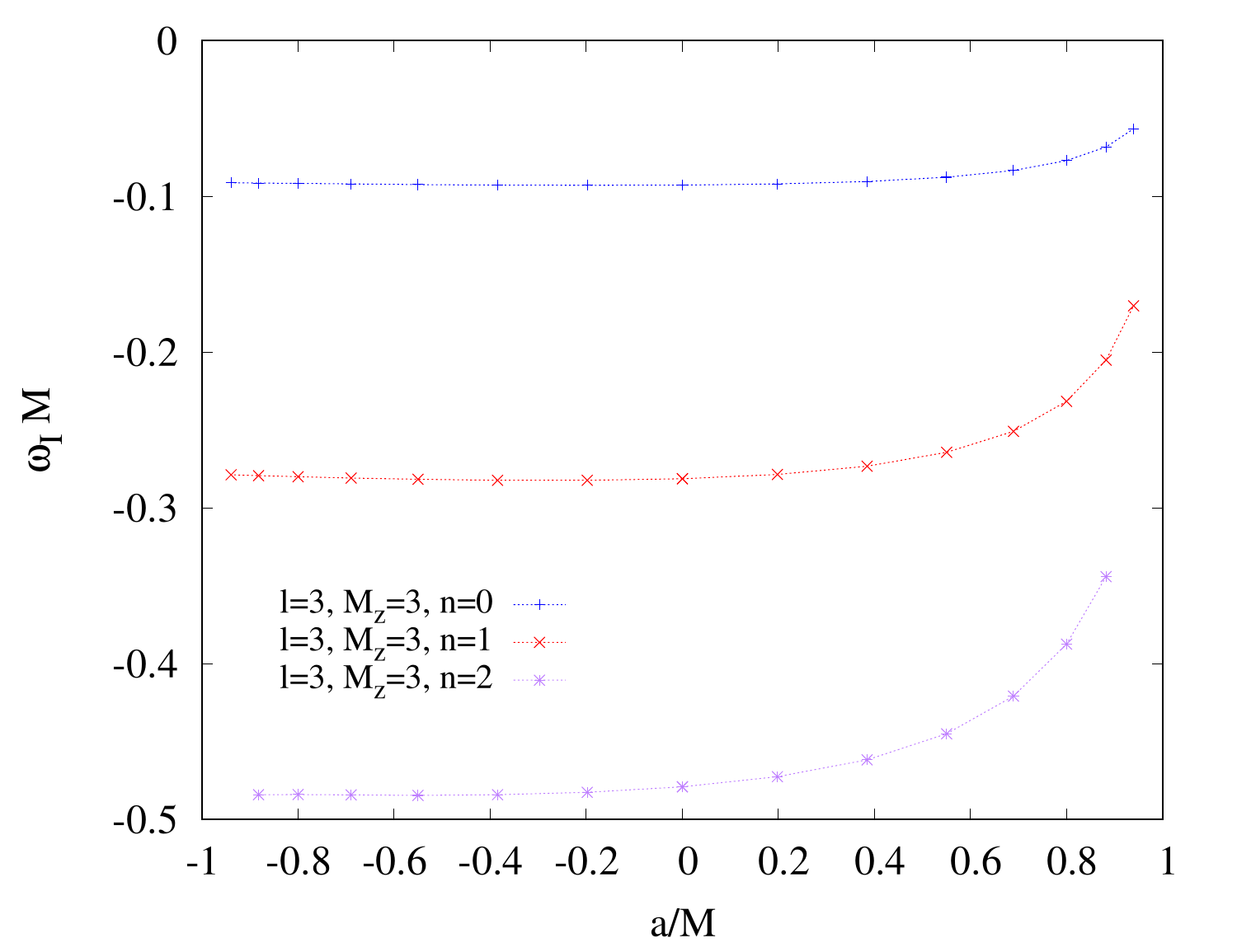}
    \caption{Quasinormal modes for $M_z=0,1,2,3$. The value of $l$ corresponds to the leading multipole of the mode in the static limit $a=0$. The value of $n$ corresponds to the excitation, where $n=0$ is the fundamental mode, i.e. the longest lived mode.}
    \label{fig:qnms_m0m1m2m3l2l3}
\end{figure}

In order to study more systematically the quasinormal modes as functions of the angular momentum, we have generated the spectrum for different values of the Kerr parameter $a$ considering $M_z=0,1,2,3$, and extracted the $(l=2)$-led and $(l=3)$-led modes. 
We also extract the $n=1,2$ excited modes, when it is possible to do so with sufficient accuracy. 
In the following, we show the results for an $N_x=N_y=20$ grid. 
These are the modes that are astrophysically interesting
since they could be part of the spectrum of the ringdown phase of the gravitational waves, although this may change for other models of gravity.

In Figure \ref{fig:qnms_m0m1m2m3l2l3} we show in the left panels $\omega_R M$ as a function of $a/M$, and in the right panels $\omega_I M$ as a function of $a/M$. 
The first row of figures is for $M_z=0$, the second for $M_z=1$, the third for $M_z=2$ and the last for $M_z=3$. 
The $l=2$ modes are shown by solid lines, and the $l=3$ modes by dashed lines.
Note that in the $M_z=3$ figures, there is obviously no $(l=2)$-led mode. 
In blue we show the fundamental mode, in red the first excitation and in purple the second excitation.

\begin{figure}
    \centering
    \includegraphics[angle=-90,width=0.45\textwidth]{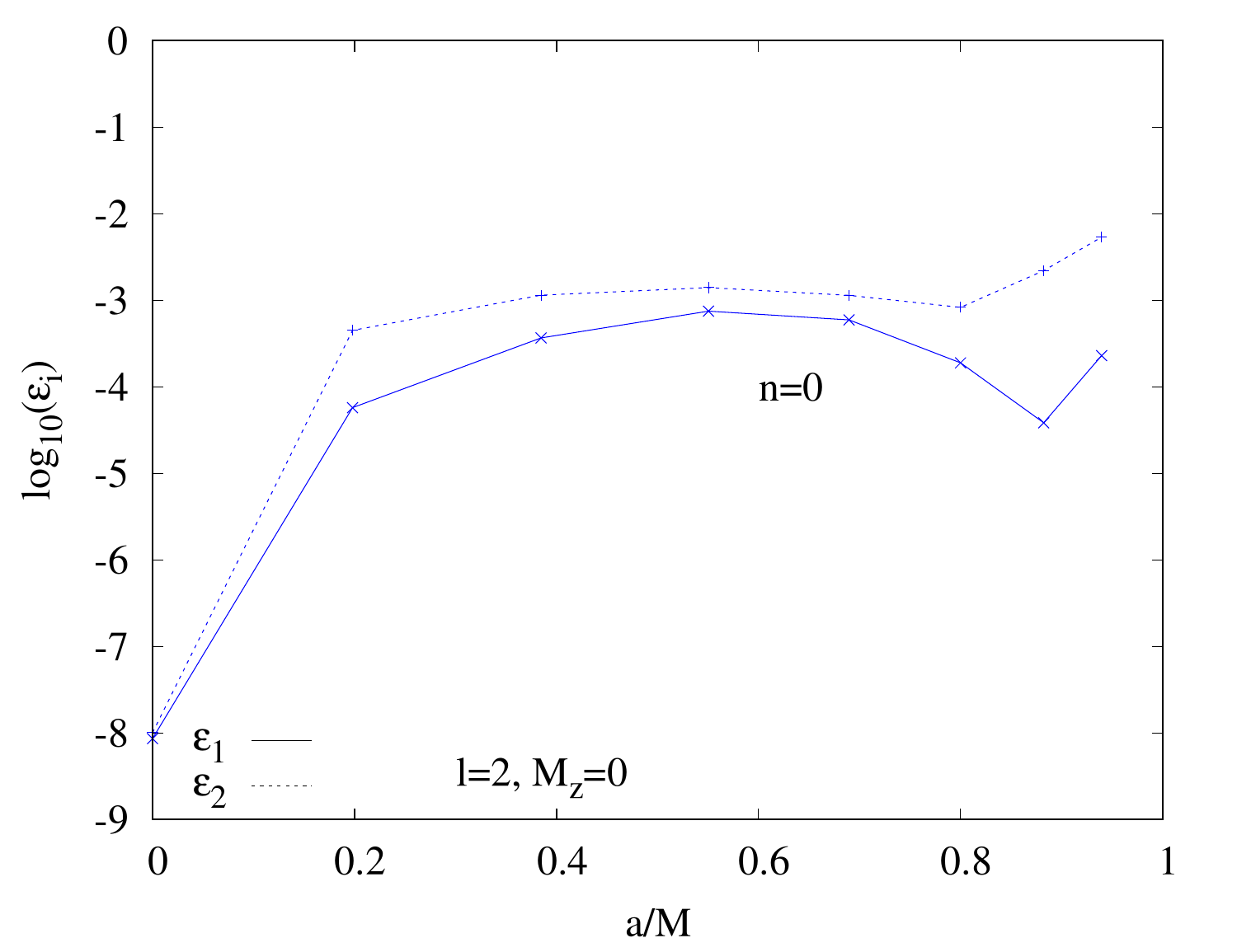}
    \includegraphics[angle=-90,width=0.45\textwidth]{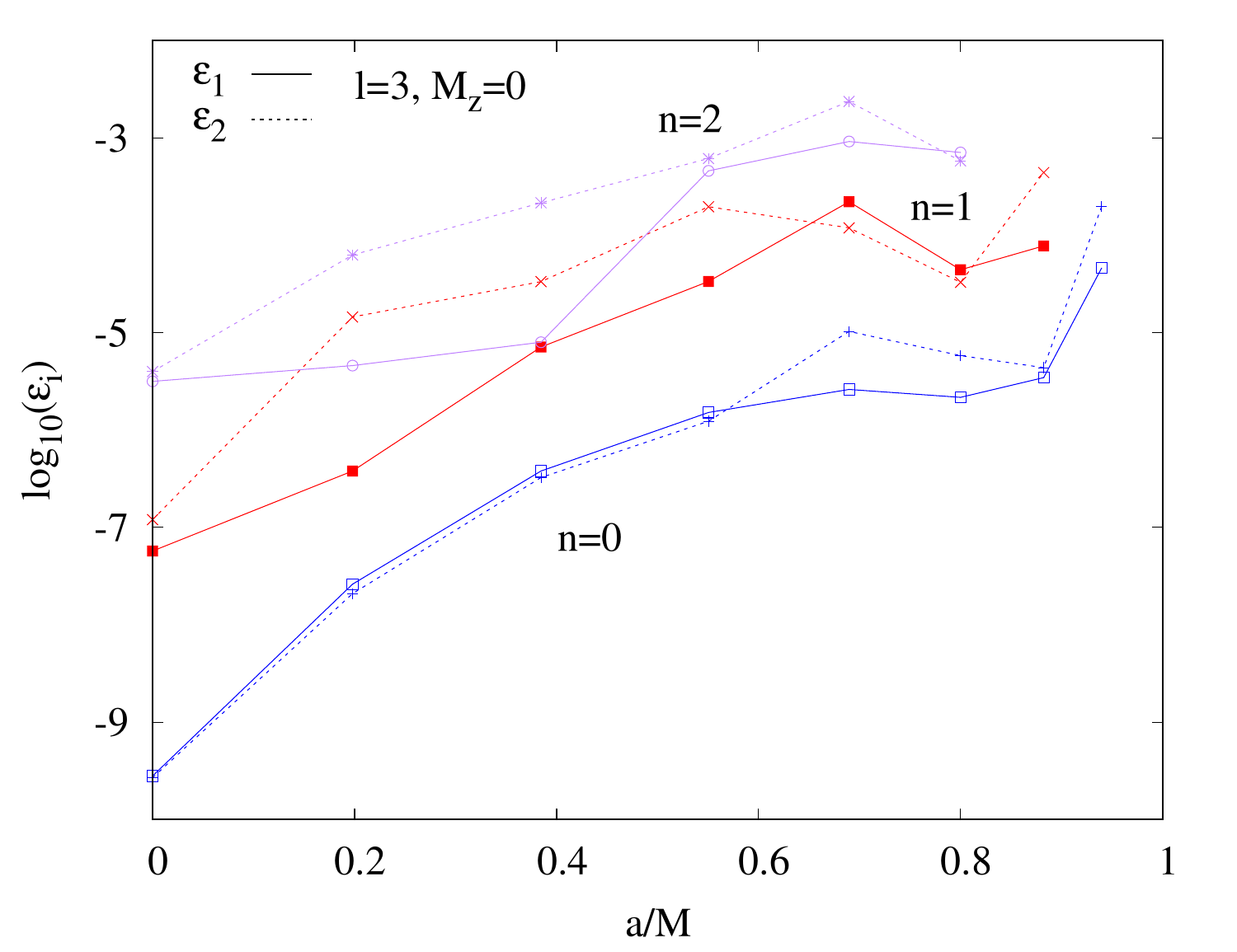}
    \includegraphics[angle=-90,width=0.45\textwidth]{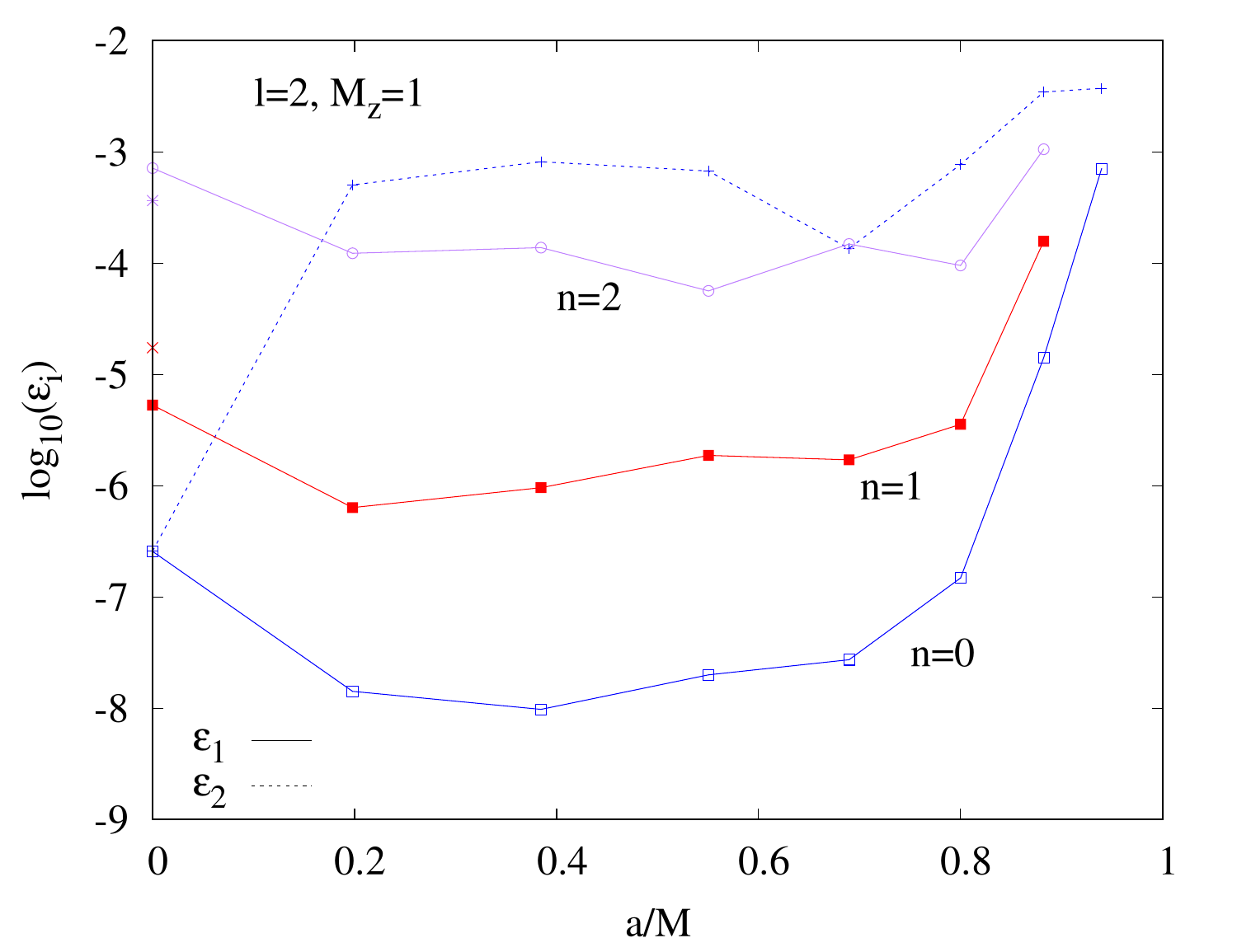}
    \includegraphics[angle=-90,width=0.45\textwidth]{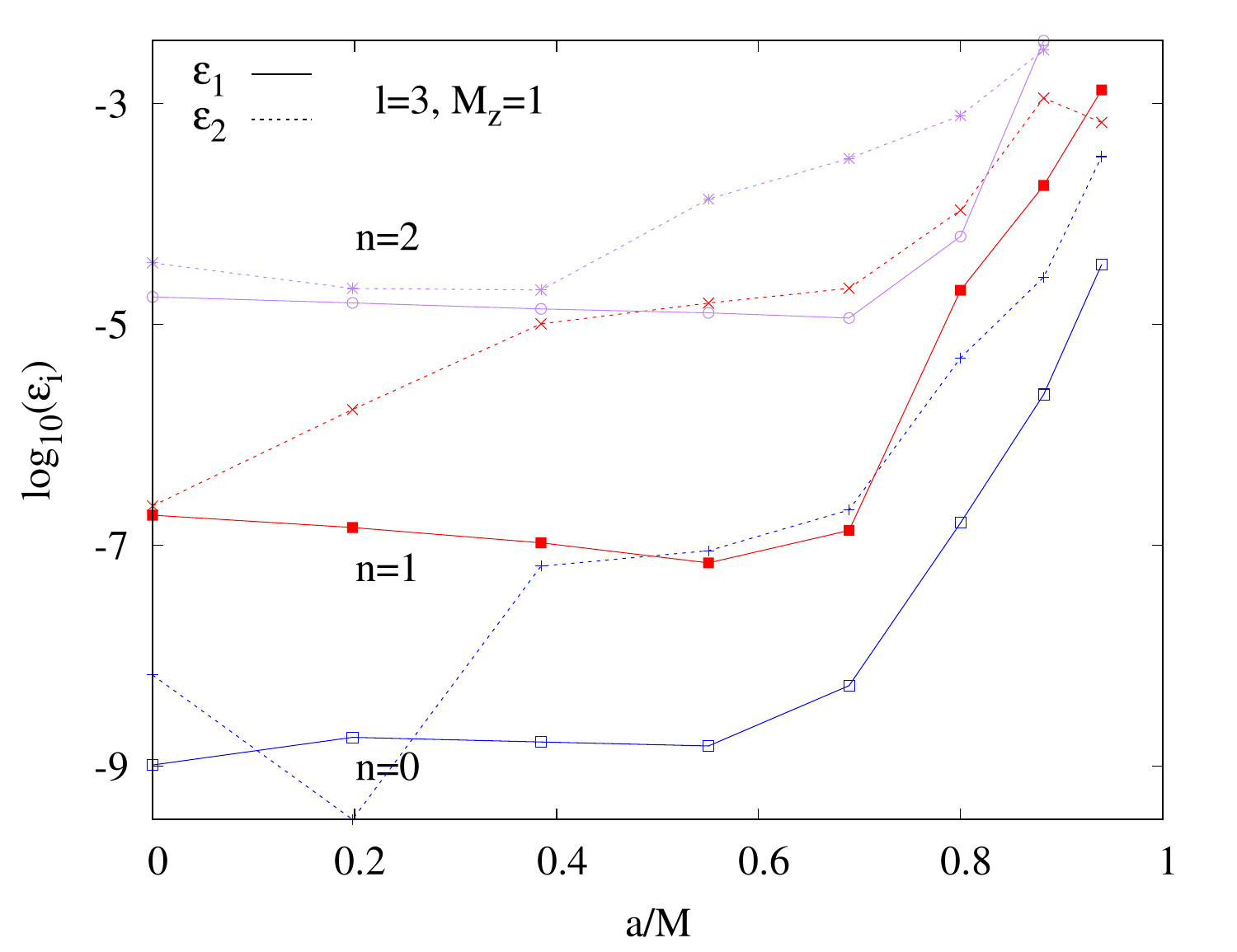}
    \includegraphics[angle=-90,width=0.45\textwidth]{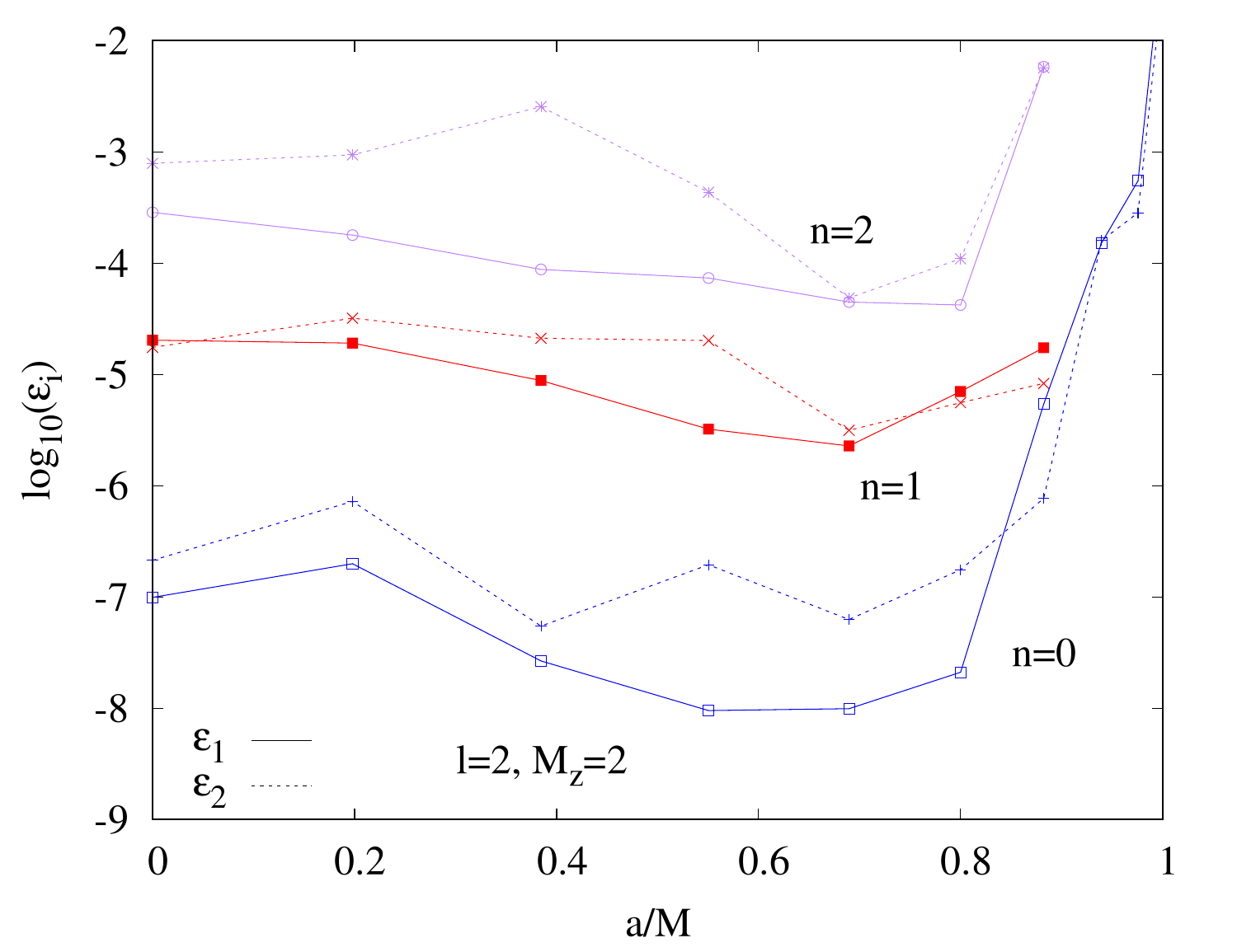}
    \includegraphics[angle=-90,width=0.45\textwidth]{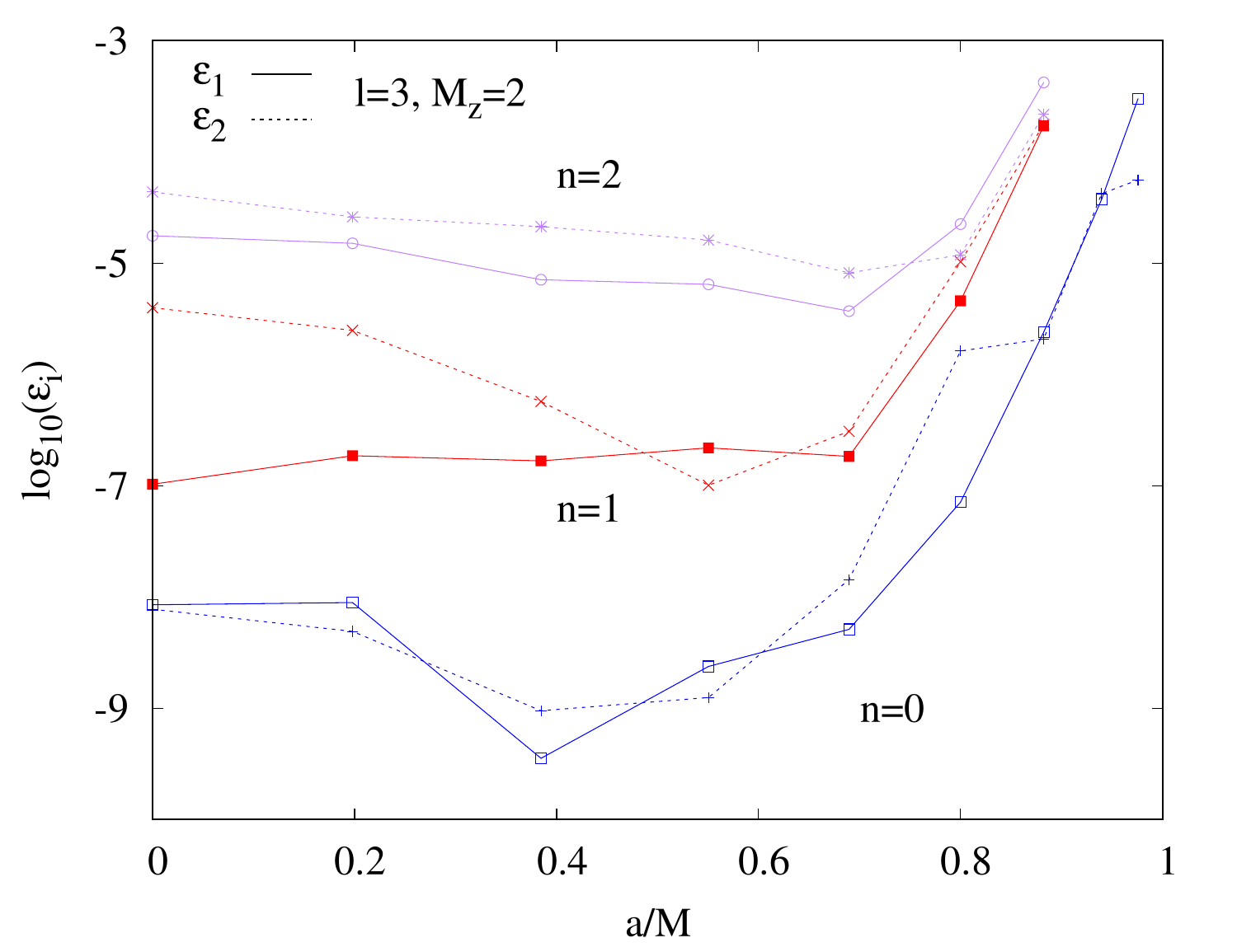}
    \includegraphics[angle=-90,width=0.45\textwidth]{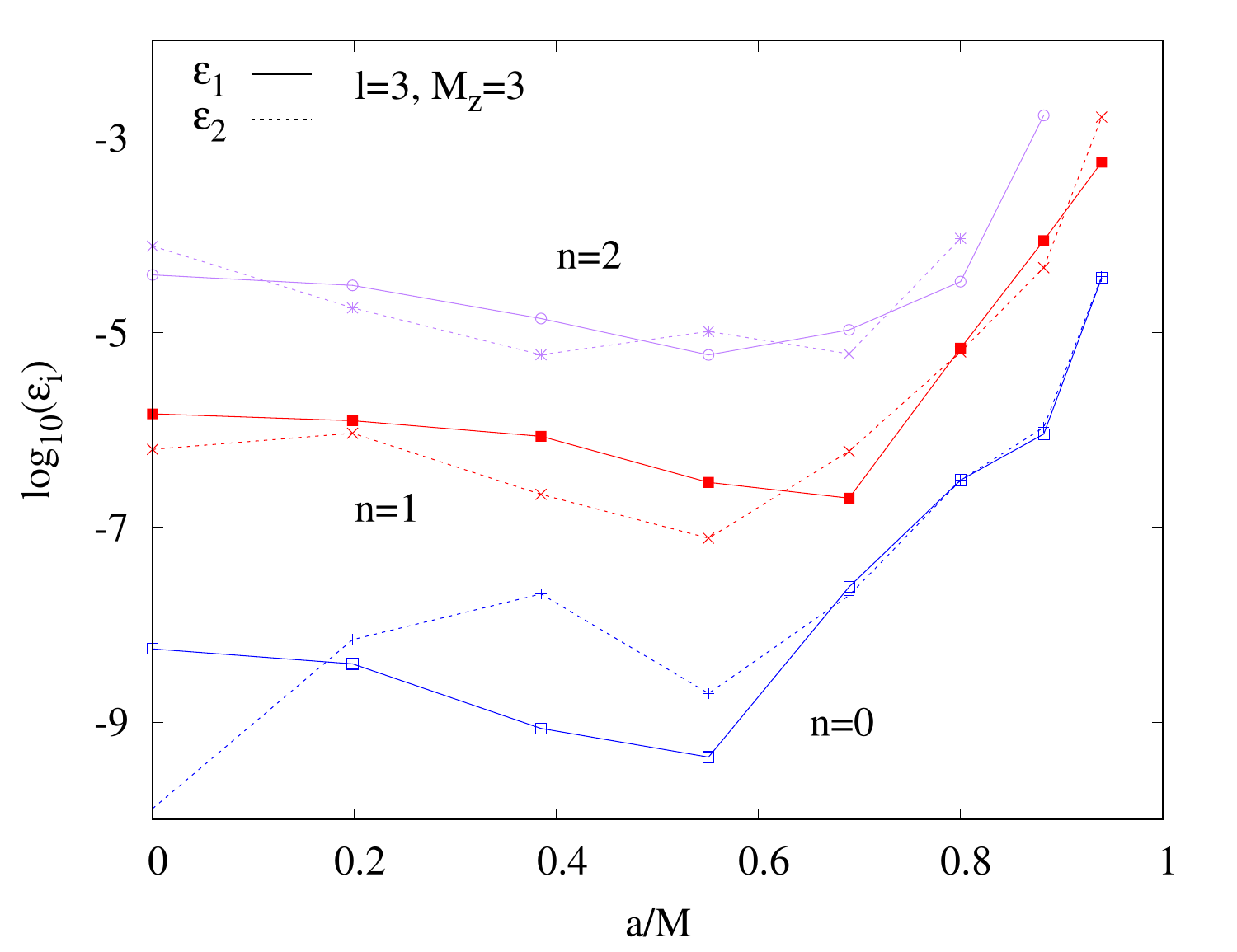}
    \caption{The estimated errors $\epsilon_1$ and $\epsilon_{2}$ in a logarithmic scale as a function of $a/M$ for Kerr quasinormal modes.}
    \label{fig:err_m0m1m2m3l2l3}
\end{figure}

In order to estimate the accuracy of our method, we compare our results with those obtained from the Teukolsky equation \cite{Berti:2009kk}. To that end we define the following quantities
\begin{eqnarray}
   \epsilon_1=|1-\omega^{(P)}/\omega^{(T)}| \, , 
   \label{with_Teu}
   \\
   \epsilon_{2}=|1-\omega^{(P)}/\omega^{(A)}| \, .
   \label{with_iso}
\end{eqnarray}
With equation (\ref{with_Teu}) we calculate the difference between our polar-led mode $\omega^{(P)}$ and the corresponding mode calculated using the Teukolsky equation $\omega^{(T)}$. With equation (\ref{with_iso}) we calculate the difference between our axial-led mode $\omega^{(A)}$ and our polar-led mode $\omega^{(P)}$.

We show in Figure \ref{fig:err_m0m1m2m3l2l3} the value of these two parameters $\epsilon_i$ for $i=1,2$ as a function of the angular momentum of the black hole. 
Note that the $y$-axis has a logarithmic scale. 
In the left panels we show the error estimations for $(l=2)$-led modes, and in the right panels for the $(l=3)$-led modes. 
The first row is for $M_z=0$, the second for $M_z=1$, the third for $M_z=2$, and the single panel at the bottom is for $l=M_z=3$. 
The curves for $\epsilon_1$ are solid, while those for $\epsilon_2$ are dashed. 
We show in blue the curves for the fundamental modes, in red and purple for the first and second excitations, respectively. 

From Figure \ref{fig:err_m0m1m2m3l2l3}, it can be appreciated that the quasinormal modes typically have an estimated relative error well below $10^{-3}$. 
The results are noticeably good for the fundamental modes in $M_z=2,3$. 
{Clearly the errors increase for faster rotating background solutions, as well as for the higher excitations of a mode. }

\begin{table}
\centering
    \begin{tabular}{|c||c|c||c|c|}
    \hline\hline
        $a/M$ & $M\omega_R^{(P)}$  & $M\omega_I^{(P)}$  & $M\omega_R^{(A)}$  & $M\omega_I^{(A)}$  \\ 
\hline\hline
0	&	0.37367166	&	-0.08896229	&	0.37367174	&	-0.08896235
\\
\hline
0.19801980	&	0.40182569	&	-0.08832052	&	0.40182598	&	-0.08832050
\\
\hline
0.38461538	&	0.43648965	&	-0.08703396	&	0.43648967	&	-0.08703397
\\
\hline
0.55045872	&	0.47837203	&	-0.08479363	&	0.47837213	&	-0.08479357
\\
\hline
0.68965517	&	0.52807220	&	-0.08117468	&	0.52807216	&	-0.08117469
\\
\hline
0.8	&	0.58601699	&	-0.07562956	&	0.58601709	&	-0.07562954
\\
\hline
0.88235294	&	0.65240125	&	-0.06754576	&	0.65240075	&	-0.06754426
\\
\hline
0.93959732	&	0.72729300	&	-0.05634091	&	0.72717684	&	-0.05637753
\\
\hline
0.97560976	&	0.81059113	&	-0.04136491	&	0.81036056	&	-0.04143177
\\
\hline
0.99447514	&	0.90265786	&	-0.00286097	&	0.91278868	&	-0.00287144
\\
\hline
\hline
    \end{tabular}
    \caption{$(l=2)$-led fundamental mode with $M_z=2$. $\omega^{(P)}$ denotes the polar-led modes, and $\omega^{(A)}$ the axial-led modes.}
    \label{tab:l2m2n0}
\end{table}

\begin{table}
\centering
    \begin{tabular}{|c||c|c||c|c|}
    \hline\hline
        $a/M$ & $M\omega_R^{(P)}$  & $M\omega_I^{(P)}$  & $M\omega_R^{(A)}$  & $M\omega_I^{(A)}$  \\ 
\hline\hline
0	&	0.34670286	&	-0.27391106	&	0.34670894	&	-0.27389876
\\
\hline
0.19801980	&	0.37862368	&	-0.27058212	&	0.37861153	&	-0.27059227
\\
\hline
0.38461538	&	0.41715522	&	-0.26531334	&	0.41716402	&	-0.26530018
\\
\hline
0.55045872	&	0.46287636	&	-0.25730339	&	0.46286703	&	-0.25729661
\\
\hline
0.68965517	&	0.51632185	&	-0.24545199	&	0.51632348	&	
-0.24544819 \\
\hline
0.8	&	0.57791808	&	-0.22814828	&	0.57792130	&	-0.22814314
\\
\hline
0.88235294	&	0.64766289	&	-0.20338646	&	0.64765752	&	-0.20332111
\\
\hline
\hline
    \end{tabular}
    \caption{$(l=2)$-led first excited mode with $M_z=2$. $\omega^{(P)}$ denotes the polar-led modes, and $\omega^{(A)}$ the axial-led modes.}
    \label{tab:l2m2n1}
\end{table}

\begin{table}
\centering
    \begin{tabular}{|c||c|c||c|c|}
    \hline
\hline
        $a/M$ & $M\omega_R^{(P)}$  & $M\omega_I^{(P)}$  & $M\omega_R^{(A)}$  & $M\omega_I^{(A)}$  \\ 
\hline
\hline
0	&	0.59944329	&	-0.09270305	&	0.59944329	&	-0.09270305
\\
\hline
0.19801980	&	0.64428239	&	-0.09198385	&	0.64428239	&	-0.09198385
\\
\hline
0.38461538	&	0.69846639	&	-0.09040666	&	0.69846641	&	-0.09040666
\\
\hline
0.55045872	&	0.76257659	&	-0.08763656	&	0.76257658	&	-0.08763656
\\
\hline
0.68965517	&	0.83698662	&	-0.08330186	&	0.83698660	&	-0.08330187
\\
\hline
0.8	&	0.92188452	&	-0.07699524	&	0.92188480	&	-0.07699528
\\
\hline
0.88235294	&	1.01730126	&	-0.06827042	&	1.01730017	&	-0.06827104
\\
\hline
0.93959732	&	1.12313403	&	-0.05661411	&	1.12309151	&	-0.05664894
\\
\hline
\hline
    \end{tabular}
    \caption{$(l=3)$-led fundamental mode with $M_z=3$. $\omega^{(P)}$ denotes the polar-led modes, and $\omega^{(A)}$ the axial-led modes.}
    \label{tab:l3m3n0}
\end{table}

\begin{table}
\centering
    \begin{tabular}{|c||c|c||c|c|}
    \hline\hline
        $a/M$ & $M\omega_R^{(P)}$  & $M\omega_I^{(P)}$  & $M\omega_R^{(A)}$  & $M\omega_I^{(A)}$  \\ 
\hline\hline
0	&	0.58264365	&	-0.28129718	&	0.58264402	&	-0.28129847
\\
\hline
0.19801980	&	0.63004044	&	-0.27855400	&	0.63004102	&	-0.27855479
\\
\hline
0.38461538	&	0.68693055	&	-0.27321978	&	0.68693040	&	-0.27321999
\\
\hline
0.55045872	&	0.75372214	&	-0.26432949	&	0.75372220	&	-0.26432966
\\
\hline
0.68965517	&	0.83061030	&	-0.25081242	&	0.83060980	&	-0.25081242
\\
\hline
0.8	&	0.91763808	&	-0.23148960	&	0.91764391	&	-0.23149054
\\
\hline
0.88235294	&	1.01469030	&	-0.20502451	&	1.01473754	&	-0.20505694
\\
\hline
0.93959732	&	1.12127807	&	-0.17028091	&	1.11944042	&	-0.17046649
\\
\hline
\hline
    \end{tabular}
    \caption{$(l=3)$-led first excited mode with $M_z=3$. $\omega^{(P)}$ denotes the polar-led modes, and $\omega^{(A)}$ the axial-led modes.}
    \label{tab:l3m3n1}
\end{table}

Finally, we end this section by presenting a number of tables with the numerical values for the quasinormal modes. 
In table \ref{tab:l2m2n0} we provide the values for the polar and axial fundamental modes for $l=M_z=2$. 
In table \ref{tab:l2m2n1} we provide the corresponding values for the first excitation. 
Similarly, tables \ref{tab:l3m3n0} and \ref{tab:l3m3n1} contain the values for the $l=M_z=3$ fundamental mode and the first excitation, respectively.

\section{Conclusions}
\label{conclusions}

In this paper we report a new method to calculate the quasinormal modes of rotating black holes and compact objects in general. We have focused the analysis on Kerr, since the spectrum of quasinormal is well-known and available for numerical comparison.

Our method {combines} two new aspects:

First, we study the standard non-radial metric perturbations, introducing at the same time both axial and polar linear perturbations on the rotating metric background. 
{This is in contrast} to the previous standard approach for studying perturbations on Kerr, which has usually been done by making use of the Newman-Penrose formalism
that results in the simple and well studied Teukolsky equation \cite{1973ApJ...185..635T}. 
However, in our approach the non-radial metric perturbations result in a complicated system of PDEs.
{While} it is non-trivial to decouple these into ordinary differential equations {in the Kerr case, this difficulty could increase tremendously,} when considering other gravity theories.

Secondly,
we have devised a spectral method to solve for the quasinormal modes of such PDE systems. 
We have decomposed the perturbation functions into a sum of Chebyshev polynomials and Legendre functions. 
Clearly, the order of the expansions can be calibrated to improve the accuracy of the calculations. 

In this paper we have tested our approach by calculating the spectrum of the Kerr black hole. 
We have shown that this method successfully produces with excellent precision a good number of quasinormal modes: 
modes with different leading multipolar behaviour, as well as excitations of the spectrum. 
The accuracy of the calculation is particularly good for the fundamental modes of the more astrophysically relevant part of the spectrum. 

In the future we plan to generalize our method to studies of quasinormal modes of black holes in alternative gravity theories, and to quasinormal modes of other compact objects.\\

Note added: While finalizing our manuscript a similar study was reported by Chung et al.~\cite{Chung:2023wkd}.

\section*{Acknowledgement}
 We gratefully acknowledge support by DFG project Ku612/18-1, 
FCT project PTDC/FIS-AST/3041/2020, COST Actions CA15117 and CA16104, and MICINN project PID2021-125617NB-I00 ``QuasiMode".
JLBS gratefully acknowledges support from Santander-UCM project PR44/21‐29910. We thank Francisco Navarro Lérida for discussions and comments on the manuscript.

\end{document}